\documentclass[twocolumn,prd,showpacs,superscriptaddress,nofootinbib,floatfix]{revtex4}
\usepackage{latexsym}
\usepackage{graphicx}

\usepackage{graphicx} 
\usepackage[usenames]{color}
\usepackage{graphicx}

\definecolor{grey}{rgb}{0.75,0.75,0.75}
\definecolor{Orange}{rgb}{1.0,0.5,0.15}
\definecolor{brown}{rgb}{0.7,0.25,0.0}
\definecolor{pink}{rgb}{1.0,0.5,0.5}
\definecolor{darkerred}{rgb}{0.8,0,0}
\definecolor{darkerblue}{rgb}{0,0,0.8}
\definecolor{Blue}{rgb}{0,0.08,0.65}
\definecolor{Red}{rgb}{0.65,0.08,0.05}
\definecolor{Green}{rgb}{0.15,0.45,0.25}

\def\red{\color{Red}}

\def\d{{\mathrm{d}}}  

\def\aeta{A\&A }
\def\aetal{A\&AL }

\def\apj{ApJ }
\def\apjs{ApJS }
\def\aj{AJ }
\def\mn{MNRAS }
\def\pasp{PASP }
\def\apjl{ApJL \rm}

\def\mnras{MNRAS }

\def\iint{\int\!\!\!\!\!\int}    % double integral
\def\iiint{\iint\!\!\!\!\!\int}  % triple integral
\def\mathfrak#1{{\mathrm{#1}}}% roman font in math mode
\def\R#1{{\mathrm{#1}}}% roman font in math mode
\def\Sec#1{{Section~\ref{s:#1}}}
\def\Eq#1{{Eq.~\ref{e:#1}}}% equation reference
\def\Ep#1{{~(\ref{e:#1})}}% equation reference
\def\Eqs#1#2{{equations.~(\ref{e:#1})-(\ref{e:#2})}}
\def\EQN#1{\label{e:#1}}        % eqn labelling a la Texsis
\def\Fig#1{{Fig.~\ref{f:#1}}}% figure reference
\def\Fip#1{{~\ref{f:#1}}}% figure reference

\def\M#1{{\mathbf{#1}}}% matrix notation
\def\V#1{{\mathbf{#1}}}% matrix notation
\def\M#1{{\mathbf #1}}

\def\bfr {{\bf r}}

\def\ba {{\mathbf a}}

  \def\br{{\bf r}}

   \def\b1{{\bf 1}}

\def\be{{\bf e}}

\def\ie{{\frenchspacing\it i.e. }}

\def\R#1{{\mathrm{#1}}}         % roman font in math mode
\def\Eq#1{{Eq.~(\ref{e:#1})}}   % equation reference
\def\Ep#1{{~(\ref{e:#1})}}      % equation reference
\def\Eqs#1#2{{Eqs.~(\ref{e:#1})-(\ref{e:#2})}}
\def\EQN#1{\label{e:#1}}        % eqn labelling a la Texsis
\def\Fig#1{{Fig.~(\ref{f:#1})}}

\def\Fip#1{{~(\ref{f:#1})}}
\def\be{\begin{equation}}
\def\ee{\end{equation}}
\def\ba{\begin{eqnarray}}
\def\ea{\end{eqnarray}}

\def\M#1{{\mathbf{#1}}} % matrix notation
\def\d{{\R{d}}}         % integrant
\def\Comment#1{{\red skipped text}}

      \def\br{{\bf r}}
 
   \def\r{\textsf{r}}

\def\ba{{\bf a}}

\def\ie{{\frenchspacing\it i.e. }}

\newcommand{\nicefrac}[2]{\leavevmode\kern.1em
            \raise.5ex\hbox{\the\scriptfont0 #1}\kern-.1em
      /\kern-.15em\lower.25ex\hbox{\the\scriptfont0 #2}}

\begin{document}

\title{Lattice Melting  and Rotation
in Perpetually
Pulsating Equilibria}

\author{C. Pichon$^{1}$, D. Lynden-Bell$^{1,2}$,  J. Pichon$^{3}$
 and R. Lynden-Bell$^{4}$}
\affiliation{$^{1}$  Institut d'Astrophysique de Paris UMR 7595, UPMC, 98 bis boulevard
       d'Arago, 75014 Paris, France.\\
$^{2}$ Institute of Astronomy and Clare College, Madingley Road, Cambridge CB3 0HA, United Kingdom,\\
 $^{3}$  Lyc\'ee Blaise Pascal, 20, rue Alexandre Fleming, 91400 Orsay, France,\\
 $^{4}$ 
University Chemical Laboratory,
Lensfield Road,
Cambridge CB2 1EW
England, United Kingdom.}

\begin{abstract}
Systems whose potential energies consists of pieces that scale as
$r^{-2}$ together with pieces that scale as $r^{2}$, show no violent
relaxation to Virial equilibrium but may pulsate at considerable amplitude
for ever. Despite this pulsation these systems form lattices
when the non-pulsational ``energy'' is low, and these disintegrate as that
energy is increased. The ``specific heats'' show the expected halving as
the ``solid'' is gradually replaced by the ``fluid'' of independent
particles. The forms of the lattices are described here for $N\leq 20$ and
they become hexagonal close packed for large $N$.
In the larger $N$ limit, a shell structure is formed. 
%When it is rescaled and set to rotate 
Their large $N$ behaviour is analogous to a $\gamma=5/3$ polytropic 
fluid  with a quasi-gravity such  that every 
element of fluid attracts every other in proportion to their separation.
For such a fluid, we study the ``rotating pulsating equilibria'' and 
their relaxation back to uniform but pulsating rotation. 
We also compare the rotating pulsating fluid to its discrete counter part,
and study the rate at which the rotating crystal redistributes angular 
momentum and mixes as a function of extra heat content.
\end{abstract}

\maketitle

%%%%%%%%%%%%%%%%%%%%%%%%%%%%%%%%
\section{Introduction}
%%%%%%%%%%%%%%%%%%%%%%%%%%%%%%%%
%%%%%%%%%%%%%%%%%%%%%%%%%%%%%%%%

 \begin{table*}\unitlength=0.5cm
  {\centering \begin{tabular}{ll}
  \hline
5 &  Two tetrahedra joined on a face \\
6 & Two  pyramids joined at their  bases\\
 7 & Two pentagonal pyramids joined at their bases\\
 8 & A twisted cube\\
 9 & Three pyramids, each pair sharing the edge of a base\\
 10 & Skewed pyramid twisted\\
 11 & Two skewed pyramid twisted and one center\\
 12 &Icosahedron i.e. two skewed pentagonal  pyramids\\
 13 & Icosahedron  and one  center\\
14 &  Hexagonal and pentagonal  pyramids and one center\\
15 & Two skewed hexagonal pyramids and one center\\
16 & Pentagonal pyramid and an hexagon and a triangle and one center\\
17 & Hexagonal pyramid over an hexagon over a triangle and one center\\
18 & Two pentagonal pyramids and one twisted pentagon and one center\\
\hline
 \end{tabular}\par}
 \label{table}
\caption{
First 18 configurations of equilibria. A few of them are shown on  Figure~\ref{f:fig-lown}.  Note that as the number of elements increases, the 
final configurations need not be unique. 
}
  \end{table*}

%\Xtophe{TODO: 
%1 apply predicted position of layer with new exact radii.
%2 mixing entropy for relaxation ?
%3 reproduce some of the math of paper III
%}

Astrophysics has provided several new insights into ways statistical mechanics
may be extended to cover a wider range of phenomena.
Negative heat capacity, bodies which get cooler when you heat them, were 
first encountered by \cite{edding} and when such bodies were 
treated thermodynamically by \cite{Antonov}\cite{DLBW}, what seemed natural to astronomers
was seen as an apparent contradiction in basic physics to those from 
statistical mechanics background. 
Even after a physicist \cite{thirring} first resolved this paradox 
and \cite{LB20} gave an easily soluble example,
there was considerable reluctance to accept the idea that micro canonical 
ensembles could give such different results to canonical ones.
There was still some reluctance even in 
1999 \cite{LB0}, though to those working on simulations of small clusters of atoms or
molecules, the distinction was well understood by the early 1990s, see
{\it e.g. } \cite{Bogdan}. However by 2005, the broader statistical mechanics community had embraced these ideas, and emphasized this difference 
which had been with us for 30 years (see Pichon \& Lynden-Bell 2006,
where an early account of this 
work is given).
Another area to which statistical mechanics might extend is that of collisionless systems which may be treated as ``Vlasov'' fluids in phase space.
While early attempts at finding such equilibria gave  interesting formulae reminiscent of a Fermi-Dirac distribution \cite{DBLviolent},
there is ample evidence from astronomical simulations that these equilibria
are not reached in gravitationnal systems. There are also different ways of 
doing the counting that lead to different results and recently, \cite{Arad} gave 
an example that demonstrates that the concept of a unique final state determined by a few constants of motion found from the initial state is not 
realized.  Thus the statistical mechanics of fluids in phase space remains 
poorly understood with theory and at best losely correlated with 
simulations and experiments  (but see \cite{binney}).  

This paper is concerned with a third area of non standard statistical mechanics
which is restricted to very special systems, those for which the 
oscillation of the  scale of the system separates off dynamically from the behaviour of the 
rescaled variables that are now scale free. 
These systems were found as a bi product of a study which  generalized Newton's soluble N-body problem \cite{LB2} \cite{LB3}
(papers I and II). A one dimensional model with exact solutions was found by \cite{Calogero} and the model considered here is a three dimensional generalisation of his combined with Newton's. 
A first skirmish with the statistical dynamics of
 the scale free variables despite the continuing oscillation
 of the scale was given there. Since then,  \cite{LB4}, hereafter paper III, have showed that the peculiar 
 velocities of the particles do indeed relax, as predicted, to a Maxwellian 
 distribution, whose temperature continuously changes as $ ({\rm scale})^{-2}$.
 This occurs, whatever the ratio of the relaxation time to the pulsation period.
 The peculiar velocities are larger whenever the system is smaller.
 When the system rotates at fixed angular momentum, the resulting statistical mechanics
 leads again to Maxwell's distribution function {\sl relative} to the rotating and 
 expanding frame (see paper III, equation 17).
 It is then the distribution of the peculiar velocities after the ``Hubble expansion velocity'' and 
 the time dependent rotation are removed that are distributed Maxwellianly; 
 when the system is fluid, the density distribution
 is  a Gaussian flattened according to  the rotation that results from the fixed angular
 momentum.
 
 The interest of this problem for those versed in statistical mechanics
 is that it is no longer an energy that is shared between the different
 components of the motion. The interest for astronomers lies in part because 
 these systems suffer no violent relaxation to a size that obeys the Virial 
 theorem. 
  Nevertheless, despite the continual pulsation of such systems
 the rescaled variables within them do relax to a definite equilibrium.

The form of interaction potential ensures that neither the divergence of the potential 
energy at small separations 
nor the divergence of  accessible volume at large separations 
which so plague systems with normal gravity occur for this model. Thus its interesting
different statistical mechanics is simpler and free of any controversial divergences  thanks to  the small range repulsion and the harmonic 
long range attraction  which extends to arbitrarily large separations.

Although the harmonic long range attraction is not realised in nature (with the possible exception of quarks), nevertheless, within homogeneous bodies of elipsoidal shape ordinary inverse square gravity does lead to harmonic forces analogous to those found here. Section 2.1.1 shows that only force laws with our particular scalings give such exact results. 

In this paper, we demonstrate that such systems can form solids
(albeit ones that pulsate in scale).
We study in \Sec{crystal} their behaviour in the large $N$ limits
 and 
show that they stratify into shells.

We show the phase transition  as these structures 
melt and the corresponding changes in ``specific heat''.
%Let us first formalize this system and 
%its gaseous analog in \Sec{deriv}.
We also discuss the  analogous  a $\gamma=5/3$ fluid system (corresponding to a classical white dwarf with 
an odd gravity, see below), 
 we predict their  equilibrium configuration 
 and investigate their
 properties when given some angular momentum.
\Sec{deriv} derives the basic formulae for the N-body system and its
 fluid analog.

%%%%%%%%%%%%%%%%%%%%%%%%%%%%%%%%
\section{Derivation}
\label{s:deriv}
%%%%%%%%%%%%%%%%%%%%%%%%%%%%%%%%
%%%%%%%%%%%%%%%%%%%%%%%%%%%%%%%%

%%%%%%%%%%%%%%%%%%%%
\subsection{The discrete system }
\label{s:nbody}
%%%%%%%%%%%%%%%%%%%%
%%%%%%%%%%%%%%%%%%%%

\begin{figure*}\unitlength=0.5cm
 {\centering \begin{tabular}{c c c}
\resizebox*{0.25000\textwidth}{0.20000\textheight}{\includegraphics{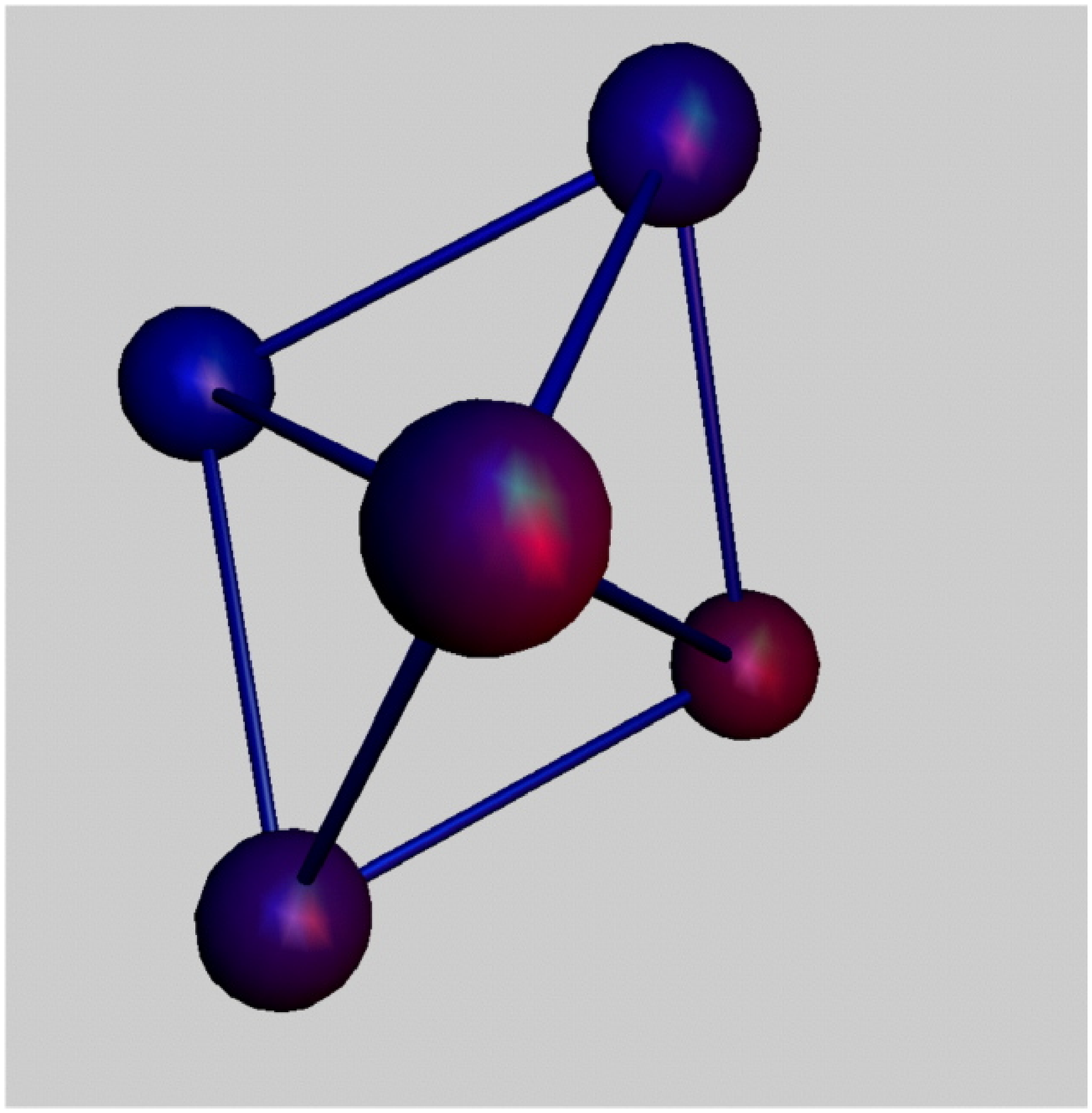}}  & 
\resizebox*{0.25000\textwidth}{0.20000\textheight}{\includegraphics{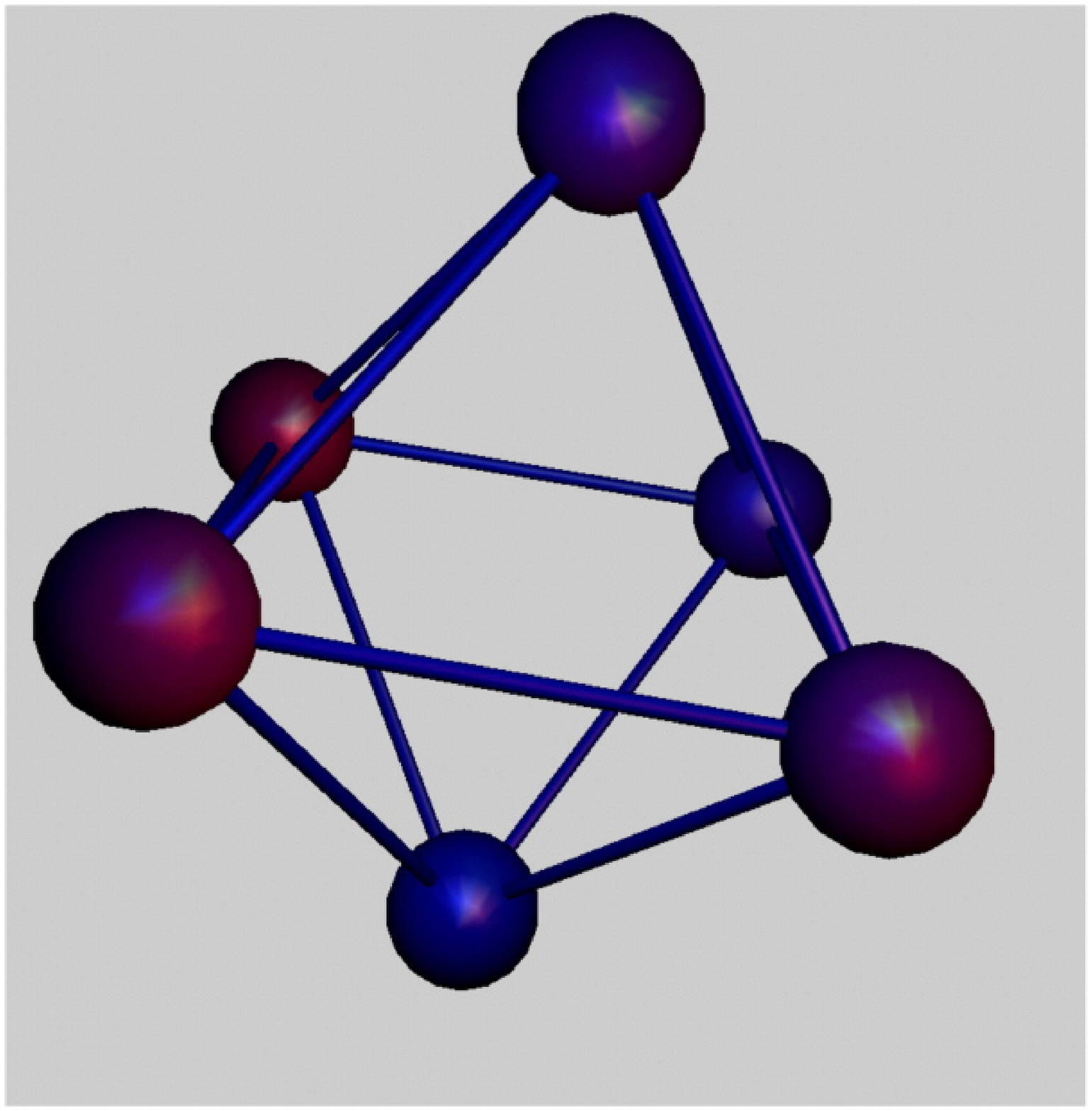}}  &  
\resizebox*{0.25000\textwidth}{0.20000\textheight}{\includegraphics{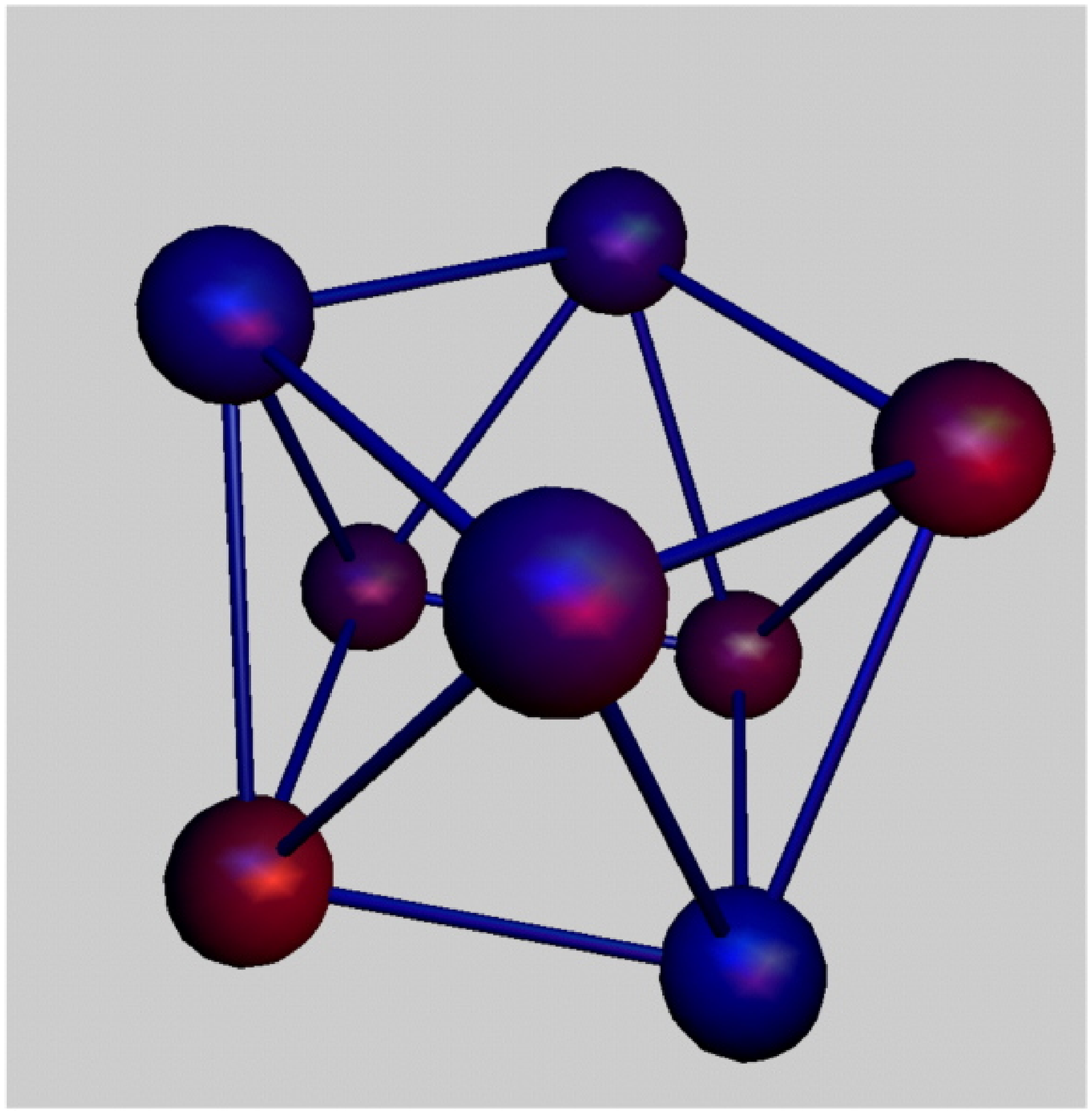}}  \\  
\resizebox*{0.25000\textwidth}{0.20000\textheight}{\includegraphics{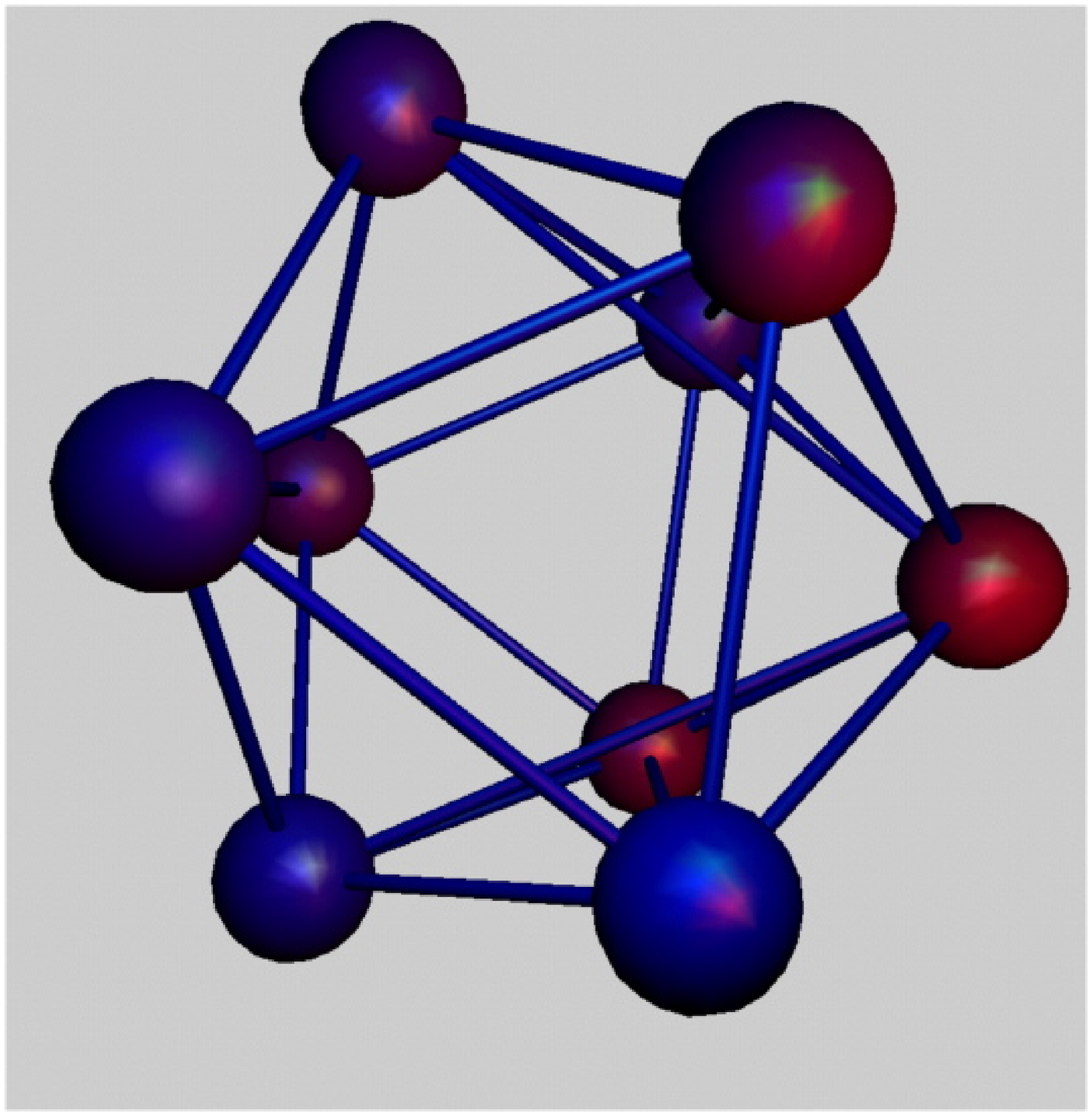}}  & 
\resizebox*{0.25000\textwidth}{0.20000\textheight}{\includegraphics{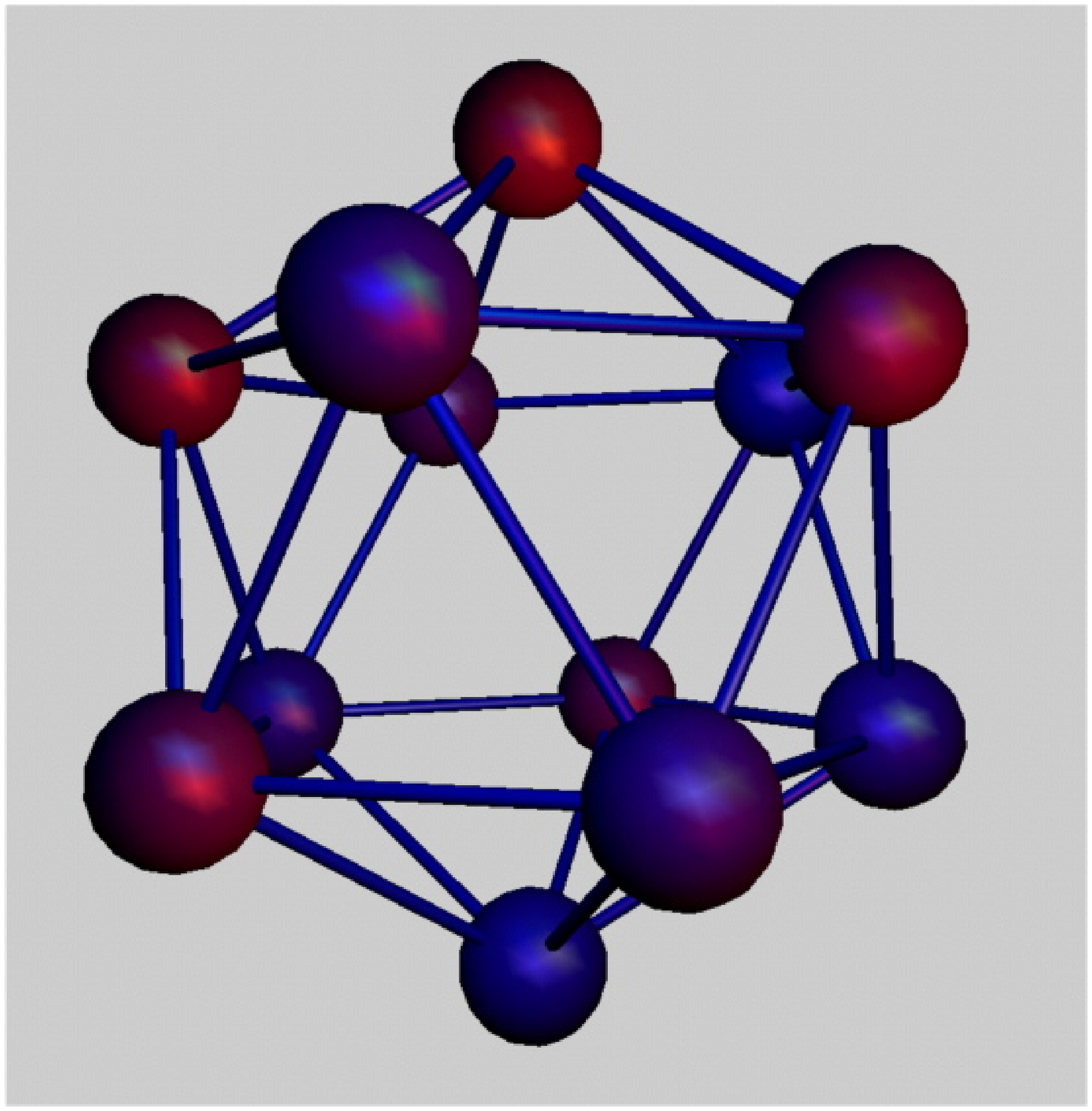}}  & 
\resizebox*{0.25000\textwidth}{0.20000\textheight}{\includegraphics{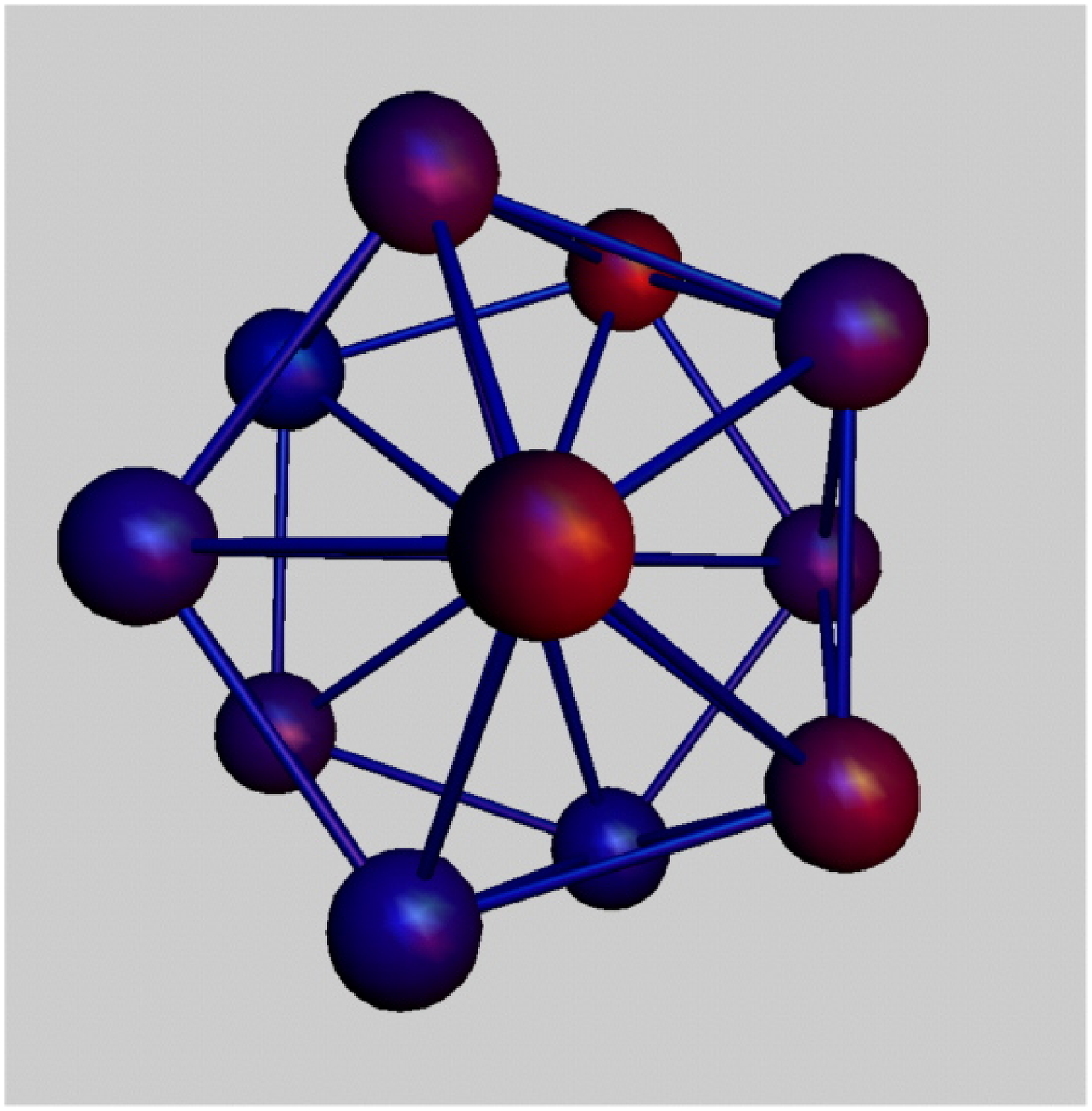}}  \\  
\resizebox*{0.2500\textwidth}{0.20000\textheight}{\includegraphics{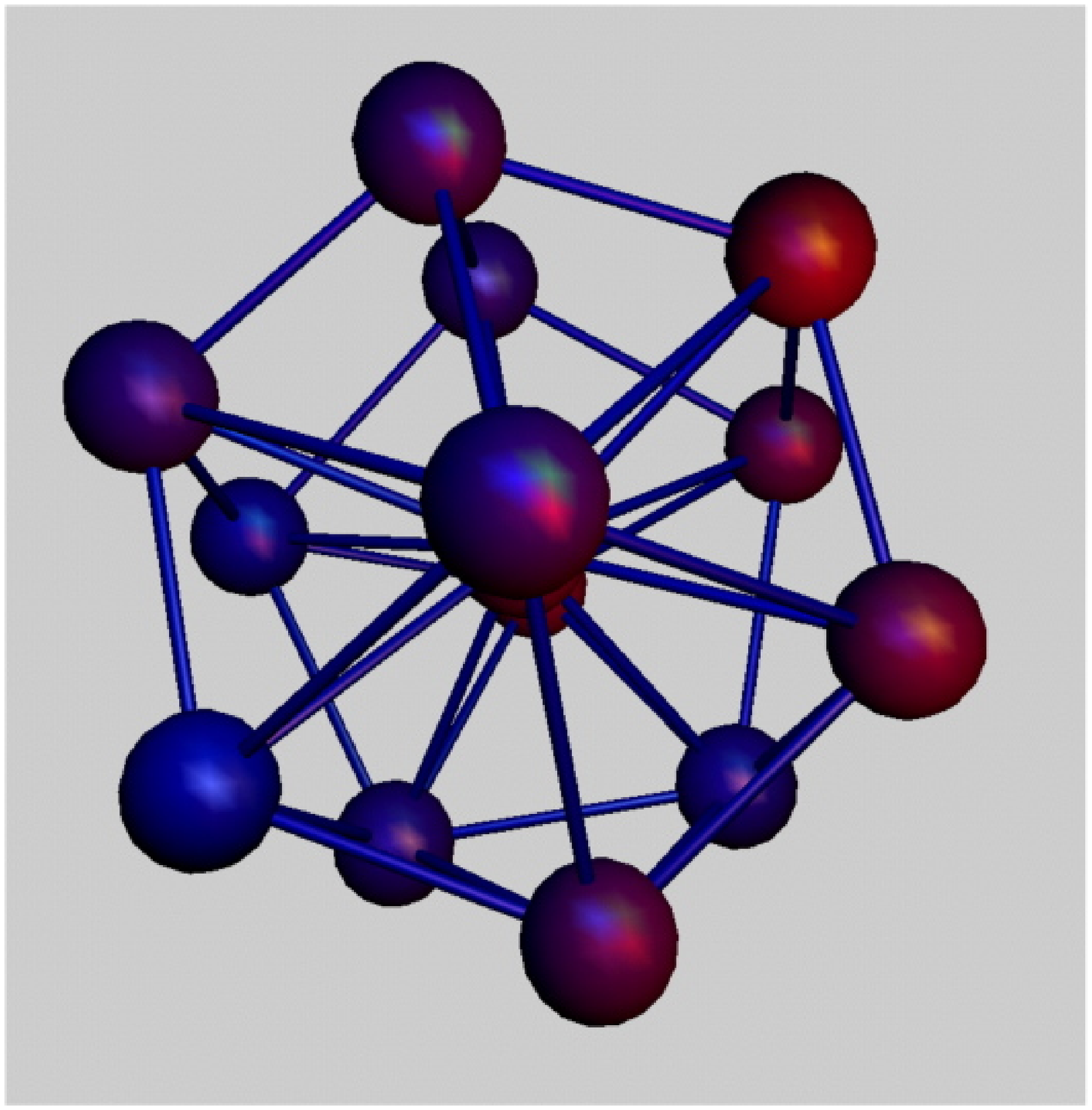}}  & 
\resizebox*{0.25000\textwidth}{0.20000\textheight}{\includegraphics{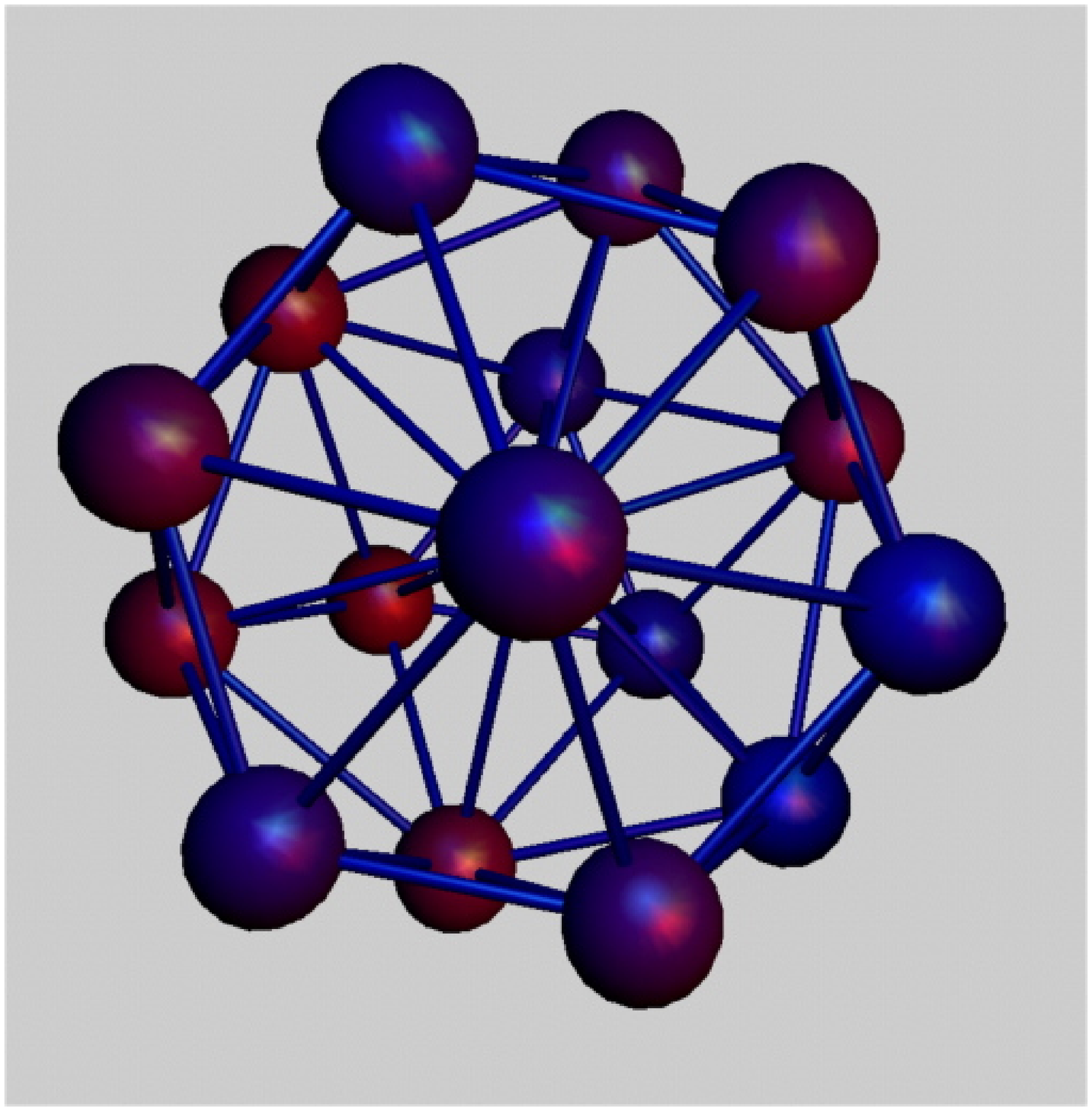}}  &  
\resizebox*{0.25000\textwidth}{0.20000\textheight}{\includegraphics{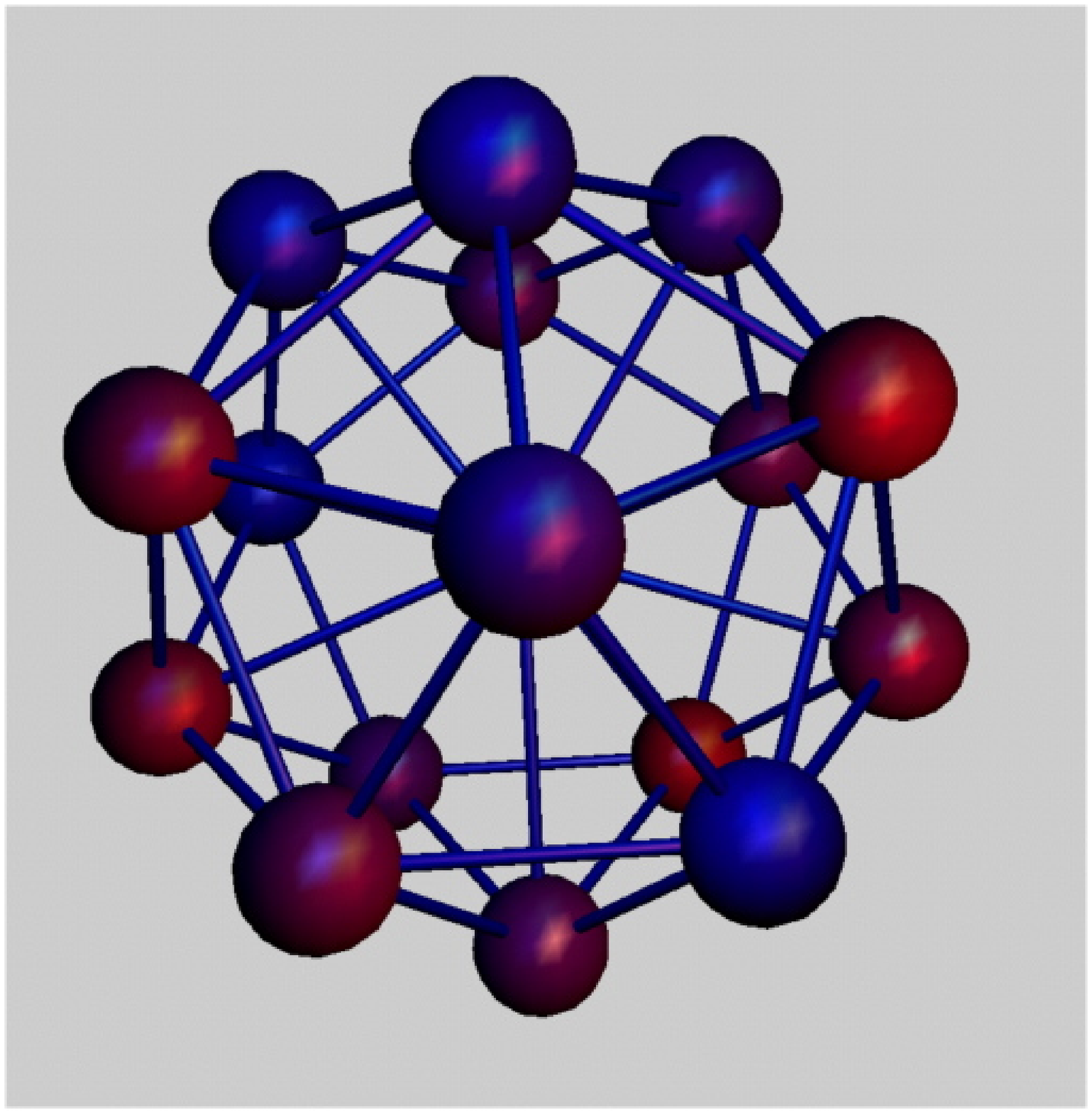}}  \\  
 \end{tabular}\par}
\caption{ (Color online)
A small set  of remarkable equilibria for low $N$; from left to right and top to bottom: $N=$
[5]   two tetrahedra joined at their bases,
[6]  two  pyramids joined at their bases,
 [8] a twisted cube (three pyramids each pair sharing a base edge),
 [12] Icosahedron i.e. two skewed pentagonal  pyramid,
 [13]  Icosahedron+center,
[14]   hexagonal+ pentagonal  pyramids+center,
[17]  hexagonal pyramid over hexagon over triangle+center,
[18] two pentagonal pyramids+twisted pentagon center.
Java animations describing theses  oscillating
crystals are found at   {\em\tt http://www.iap.fr/users/pichon/nbody.html}.
 }
\label{f:fig-lown} 
\end{figure*}

We consider $N$ particles of masses $m_i$ at ${\bf r}_i,~i=N$ and set
\begin{equation}
\sum m_i=M,~\quad \sum m_i{\bf r}_i =M\overline{\bf r}\,, \EQN{firsteq}
\end{equation}
 and half the trace of the intertial tensor as
  \begin{equation}
I=\sum
m_i(\M{r}_i-\overline{\bf r})^2\equiv{M} a^2, \EQN{defI}
\end{equation}
which defines the scale $a$.
 In earlier work we showed both classically (paper I) and quantum mechanically
 \cite{LB3}  (paper II) that if the potential energy of the whole system was of the form
\begin{equation}\label{1}
V=W(a)+a^{-2}W_2(\hat{\bf a})\,,
\end{equation}
where $\hat{\bf a}$ is the 3N dimensional unit vector
\begin{equation}
\hat{\bf a}=
\frac{N^{-1/2}}{a}({\bf r}_1-\overline{\bf r},~{\bf r}_2-\overline{\bf r}~
....~, ~{\bfr}_N-\overline{\bf r})\,,
\end{equation} then the motion of the scaling
variable $a$ separates dynamically from the motions of both
$\overline{\bf r}$ and the $\hat{\bf r}$ so those motions
decouple. If we ask that $V$ be made up of a sum of pairwise
interactions then  the hyperspherical potential, $W$, 
(independent of direction in the $3N -3$ dimensional space) has to be of the form:
\begin{equation}
W(a)=\frac{1}{2}\omega^2M a^2=V_{-2}\,,  \EQN{defW}
\end{equation}
where $\omega^2$ is constant but depends linearly on $M$, and 
\begin{equation}W_2(\hat{\bf
a})=a^{2} V_2=a^{2} \sum\sum\limits_{i<j}K_2|r_i-r_j|^{-2}\,. \EQN{defW2}
\end{equation}
Hence $W_{2}$ is a function of $\hat{\bf
a}$ and is independent of the scale $a$.  

\subsubsection{Relationship to the Virial Theorem}
\label{s:virial}
 Let us now see
why potentials of the for $V_2+V_{-2}$ are so special by looking at
the Virial Theorem and the condition of Energy conservation.
Take the more general potential energy to be a sum of pieces $V_n$
where each $V_n$ scales as $r^{-n}$ on a uniform expansion. $n$ can be
positive or negative. Then $V=\sum V_n$. For such a system the Virial
Theorem reads:
\begin{displaymath}
\frac{1}{2}\ddot{I}=2T+\sum nV_n=2E+\sum(n-2)V_n~.
\end{displaymath}
Now we already saw that $V_{-2}\propto I$ and $V_2$ clearly drops out
of the final sum because $n-2$ is zero. Thus for potentials of the
form $V=V_{-2}+V_2$
\begin{equation}
\frac{1}{2}\ddot{I}=2E-4V_{-2}=2E-4\omega^2\left(\frac{1}{2}I\right)\,, \EQN{Idotdot}
\end{equation}
so $I$ vibrates harmonically with angular frequency $2\omega$ about a
mean value $E/\omega^2$ and shows no Violent Relaxation
(1). Multiplying by $4\dot{I}$ and integrating
\begin{displaymath}
\dot{I}^2=8EI-4\omega^2I^2-4 {M}^2\mathcal{L}^2\,,
\end{displaymath}
where the last term is the constant of integration chosen in conformity with \Eq{defL}. Now recall from \Eq{defI} that
$I={M} a^2$ so we find on division by $8 M^2 a^2$ that
\begin{equation}
\frac{\dot{a}^2}{2}+\frac{\mathcal{L}^2}{2 a^2}+\frac{1}{2}\omega^2a^2=
\frac{ E}{M} \,,  \EQN{defL}
\end{equation}
which we recognise as the specific energy of a particle with specific  ``angular
momentum'' $\mathcal{L}$ moving in a simple harmonic spherical potential
of `frequency' $\omega$. In such a potential $a$ vibrates about its
mean at `frequency' $2\omega$.

Note here that the $\gamma=5/3$ fluid also has a term $3(\gamma -1) U =2 U$ in the virial 
theorem since its internal energy, $U$, scales like $r^{-2}$ so it can be likewise absorbed into the total energy term. Thus if every elementary
mass of such a fluid attracted every other with a force linearly proportional to their separation,
then that system too would pulsate eternally, as described above in \Eq{Idotdot}
(see \Sec{gas} below).

%In \cite{LB2} we suggested that despite this pulsation in scale, the rescaled
%variables should reach a statistical equilibrium giving a Maxwell
%distribution in suitably defined peculiar velocities but that the rms
%velocity would pulsate $\propto[r(t)]^{-1}$.

% In \cite{LB3} we simulated this
%system and demonstrated that this pulsating Maxwellian distribution is
%indeed achieved. However although we noted that the potential had both
%a long range attraction and a short range repulsion suitable for the
%formation of a `solid' lattice (which must be able to pulsate at large
%amplitude because of the special nature of the system) we did not
%demonstrate such a lattice.

The aim of this paper is to demonstrate the existence of perpetually
pulsating equilibrium lattices and to study the changes as the
non-pulsational `energy' ${M}\mathcal{L}^2/2$ is
increased. We show that the part of the potential left in the equation
of motion for the rescaled variables is the purely repulsive $V_2$. It
is then not surprising that the lattice disruption at higher
non-pulsational energy occurs quite smoothly %without any latent heat
and the solid phase appears to give way to the ``gaseous'' phase of half
the specific heat without the appearance of a liquid with another
phase transition. The hard sphere solid has been well studied and
behaves rather similarly.

\begin{figure}\unitlength=0.5cm
 {\centering
 \resizebox*{0.450000\textwidth}{0.450000\textwidth}{\includegraphics{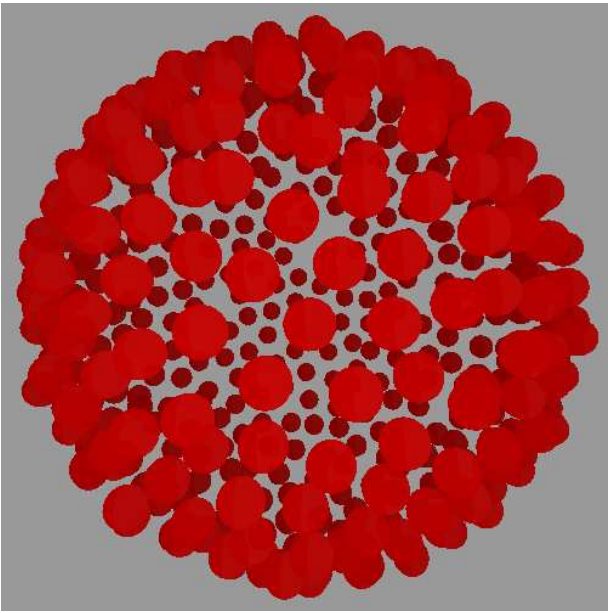}} 
 \par}
\caption{ (Color 
  online)
An example of large $N$ static spherical equilibrium obeying Eq.~(\ref{2}); here only a given shell is represented
for clarity.
}
\label{f:3dshell}
\end{figure}

\subsubsection{The Equations of Motion and their separation}
%%%%%%%%%%%%%%%%%%%%%%%%%%%%

Although the work of this section can be carried out when the masses
$m_i$ are different, (see paper I), we here save writing by taking $m_i=m$
and so $M=Nm$. We start with $V$ of the form (\ref{1}). The equations
of motion are
\begin{equation} \label{2}
m\ddot{\bf r}_i =-\partial V/\partial {\bf r}_i~~.
\end{equation}
Now $V$ is a mutual potential energy involving only ${\bf r}_i-{\bf
r}_j$ so $\sum\limits_{i}\partial V/\partial{\bf r}_i=0$ and summing
the above equation for all $i$ we have
\begin{displaymath}
d^2\overline{\bf r}/dt^2=0~~~,~~~\overline{\bf r}=\overline{\bf r}_0
+{\bf u}t~~.
\end{displaymath}
Henceforth we shall remove the centre of mass motion and fix the
centre of mass at the origin so $\overline{\bf r}=0$. Now the
$3N$-vector ${\bf a}$ can be rewritten in terms of its length $a$ and
its direction $\hat{\bf a}\equiv {\bf a}/a $ (a unit vector). Equation (\ref{2}) takes
the form 
\begin{equation} \label{3}
M\ddot{\bf a}=-\partial V/\partial{\bf a}=- W^\prime\hat{\bf
a} +2a^{-3}W_2\hat{\bf a} - a^{-3}\partial W_2/\partial\hat{\bf a}\,,
\end{equation}
where we have used the form (\ref{1}) for $V$. Notice that when $W_2$
is zero (or negligibly weak) all the hyperangular momenta of the form
$m(r_\alpha \dot{r}_\beta -r_\beta \dot{r}_\alpha)$ where $\alpha\neq\beta$
run from 1 to $3N$, are conserved!

Taking the dot product of (\ref{3}) with ${\bf a}$ eliminates the
$\partial W_2/\partial\hat{\bf a}$ term which is purely transverse so
we get the Virial Theorem in the form:
\begin{equation}
\frac{1}{2} \frac{\d{}^2}{\d t^2}(M a^2) =M \dot{\bf a}^2-a\frac{\d W}{\d a} -\frac{2}{a^{2}}W_2
=2E-\frac{1}{a} \frac{\d}{\d a}(a^2W)~. \EQN{vir1}
\end{equation}
Multiplying by $ {\d(a^2)}/{\d t}/2$ and integrating \Eq{vir1} yields
\begin{displaymath}
\frac{1}{2} M a^2 \dot{a}^2 =Ea^2 -W a^2-\frac{1}{2}M \mathcal{L}^2\,,
\end{displaymath}
where the final term is the constant of integration.  On division by
$a^2$
\begin{equation} \label{4}
\frac{M}{2}\left({\dot{a}^2}+\frac{{\mathcal L}^2}{a^2}\right) + W(a)=E\,,
\end{equation}
which is the energy equation of a particle of mass $M$ moving with
angular momentum $\mathcal{L}$ and energy $E$ in a hyperspherical
potential $W(a)$. Now 
\begin{equation}\ddot{\bf a}=\frac{\d{}}{\d t}(\dot{a}\hat{\bf
a}+{ a}\dot{\bf\hat{a}}) =\ddot{a}\hat{\bf a}+
\frac{1}{a}\frac{\d{}}{\d t}(a^2\dot{\bf\hat{a}})\,
\end{equation}
 and from (\ref{4})
\begin{equation} \label{6}
\ddot{a}=\frac{\mathcal{L}^2}{a^3}-\frac{1}{M}W^\prime(a)~~.
\end{equation}
Inserting these values for $\ddot{\bf a}$ and $\ddot{a}$ into equation
(\ref{3}) we obtain on simplification, multiplying by $a^3$:
\begin{equation} \label{7}
M a^2\frac{\d{}}{\d t}\left(a^2 \frac{\d\hat{\bf a}}{\d t}\right) = 2W_2
\hat{\bf a} - \partial W_2/\partial\hat{\bf a} -
M\mathcal{L}^2\hat{\bf a}~~.
\end{equation}
On writing $\d{}/\d\tau = a^2 \d{}/\d t$ this becomes an autonomous equation for
$\hat{\bf a}(\tau)$.
Since $(\dot{\bf a})^2 =(a\dot{\hat{\bf a}} + \dot{a}\hat{\bf a})^2
=a^2\dot{\hat{\bf a}}^2 +\dot{a}^2$ we may write the energy in the
form
\begin{displaymath}
E=\frac{1}{2} M \left[\frac{1}{a^2}\left(\frac{d\hat{\bf
a}}{d\tau}\right)^2 +\dot{a}^2\right] + V=
\frac{1}{2} M \left(\dot{a}^2 +\frac{\mathcal{L}^2}{a^2}\right)+W\,,
\end{displaymath}
so
\begin{equation} \label{8}
\frac{1}{2}M \left(\frac{\d\hat{\bf a}}{\d\tau}\right)^2 + W_2
=\frac{1}{2}M \mathcal{L}^2~~.
\end{equation}
This shows that the only `potential' in the hyper angular coordinates'
motion is $W_2(\hat{\bf a})$ and that the effective hyper  angular energy in
that motion is $\frac{1}{2}M \mathcal{L}^2$ (which has the dimension of
$a^2$ times an energy). In fact we showed in papers I and III
that it was this quantity that was equally shared among the
hyperangular momenta in the statistical mechanics of perpetually
pulsating systems. % Subtracting the radial part of
 (\ref{7}) 
gives the equation of motion for $\hat{\bf a}$. 
%The radial part
%reduces to the ${d}/{d\tau}$ derivative of equation (\ref{8}).
%
In terms of the true velocities ${\bf v}_1,~{\bf v}_2$ etc
\begin{displaymath}
\frac{\d\hat{\bf a}}{\d \tau} =\sqrt{N} \! a^2 \frac{\d{}}{\d t} \left(\frac{{\bf
r}_1}{a},~\frac{{\bf r}_2}{a}~....~\frac{{\bf r}_N}{a} \right) 
=
\end{displaymath}
\begin{displaymath}
a\left({\bf v}_1 -\frac{\dot{a}}{a}{\bf r}_1,{\bf v}_2
-\frac{\dot{a}}{a}{\bf r}_2~...~{\rm \sl etc}\right)\,,
\end{displaymath}
so it is the peculiar velocities after removal of the Hubble flow
${\dot{a}} {\bf r}_i/{a} $ and after multiplication by $a$ that
constitute the kinetic components of the shared angular energy in the
angular potential $W_2(\hat{\bf a})$.

When  the  $m (\d\hat{\bf a}/\d\tau)^2/2$ are  large enough  to
escape the potential  wells offered by $W_2/N$ we get  an almost free particle
angular motion of these $3N-4$  components. The $-4$ accounts for the fixing
of  the  centre of  mass  and  the removal  of  $\dot{a}$  from the  kinetic
components. If we impose also a prescribed total angular momentum, the number
of independent kinetic  components would reduce by a  further three
components  to $3N-7$.  However  the peculiar  velocities are  then measured
relative to a frame rotating with angular velocity ${\bf \Omega}
$  where   ${\bf  \Omega}={\bf \cal I}^{-1}\cdot {\bf J}$ where 
$\M{J}$ is the angular momentum of the system. Here $\cal I$ is the inertial 
tensor, not to be confused with $I= {\rm trace}({\cal I})/2$ introduced in 
\Sec{virial}. 
   The  constant  Lagrange
multiplier is no longer ${\bf \Omega}$,  which  is proportional to $a^{-2}$
since  the inertial tensor, has that dependence in pulsating  equilibria. The Lagrange
multiplier is ${\tilde {\bf J}}={\bf  \Omega}a^2$ which is constant during
the pulsation  and the peculiar  velocity is ${\bf v}_{\rm p  i}=({\bf v}_i-{\bf
  u}_i)$  where  ${\bf  u}_i  =  \left({\dot{a}}{\bf  r}_i/{a}+{\tilde {\bf
    J}}\times {\bf r}_i/{a^2}\right)$
       see paper III equation (17).\footnote{
The above definition of ${\bf u}$ is correct. That given under equation (18)
of paper III has $-{\bf \Omega}$ for ${\bf \Omega}$ in error as may be seen
from equation (17).}
Hereafter we specialise to the requirements given 
by \Eq{defW} and \Ep{defW2}.
% $W(r)=V_{-2}=\frac{1}{2}M \omega^2r^2$ and
%$r^{-2}W_2=V_2=\sum\sum\limits_{i<j}K_2|{\bf r}_i - {\bf r}_j|^{-2}$ as in the introduction.

%%%%%%%%%%
\subsection{The $\gamma=5/3$ analogous fluid system}
\label{s:gas}
%%%%%%%%%%%%%%

When particles attract with both a long range force such as gravity 
or our linear law of attraction, and a short rage repulsion, the latter
acts like the pressure of a  fluid. 
Indeed 
short range repulsion forces only extend over a local region,
and for large $N$ their effect can be considered as pressure since only 
the particles close to any surface drawn through the configuration 
affect the exchange of momentum across that  surface.
In our case, the local $r_{ij}^{-3}$ forces come from  the  $r_{ij}^{-2}$ potential
which scales as $r^{-2}$. 
%It turns out that only a $\gamma=5/3$ gas has the same 
%scaling with density as a $1/r^{3}$ repulsion between particle.
%The local repulsive potential scales like $L^{-2}$ and its comparatively local character suggests that is may
%be approximated  by a pressure. 
%
A barotropic fluid with $p=\rho^{\gamma}$ has an internal energy that behaves 
as $\rho^{\gamma -1}$ scaling like $r^{-3(\gamma -1)}$. The required $r^{-2}$ scaling gives a $\gamma$ of $5/3$,
\ie a polytropic index of $n=3/2$, the same as a non relativistic degenerate white dwarf.
Thus we may expect {\sl analogies} between a $\gamma=5/3$
fluid with the linear long range attraction,  and our large $N$ particle systems.
However an $r_{ij}^{-3}$ repulsion between particles is not of very short
range so this  fluid is not the same as the large $N$
limit of the particle system.
The particle equilibria have the inverse cubic repulsion between particles balancing the long range 
linear attraction, which can be  exactly replaced by a linear  attraction to the barycentre proportional to the total mass. For the fluid, it is the pressure gradient that balances this long range force.

\subsubsection{Mass profile of the  static  polytropic fluid}
%%%%%%%%%%%%%%

The equilibrium of such a fluid  in the presence of a long range force, $- G^{\star}M r $, (where $M G^{\star}\equiv \omega^{2} $)  is given by 
\begin{equation}
\frac{1}{\rho} \frac{\d p}{\d r}=- G^{\star} M r \,.\EQN{defpressprof0}
\end{equation}
Setting $p=\kappa \rho^{5/3}$,  with $\xi\equiv r/r_{\rm m}$, this yields
\begin{equation}
\rho = \rho_{0} (1-\xi^{2})^{{3/2}},\EQN{defpressprof}
\end{equation}
where $\M{r}$ is  the 3D configuration space vector
 measured from the centre of mass of the system and 
 $r$ its modulus, with $r_{\rm m}$ its value at the edge.
 
 Integrating the density gives a mass profile, $M(\xi)$,
\begin{equation}
M(\xi)= \frac{  2 M}{3 \pi}\left( 3 \arcsin(\xi)-\xi ( 8 \xi^{4} -14 \xi^{2} +3) \sqrt{1-\xi^{2}}
\right), \EQN{defmassprof}
\end{equation}
and $M$ is the total mass. In practice for the discrete system
of \Sec{nbody}, there is a marked layering at equilibrium, and the ``pressure'' and the mass profiles 
depart from their predicted profiles as each monolayer of particles is crossed.
See \Sec{crystal} and \Fig{profile-section}.

The relationship between the polytropic coefficient, $\kappa$, and $K_{2}$, the strength of 
the repulsion (entering $W_{2}$ in  \Eq{defW2}) is found by identifying the 
internal energy of the corresponding fluid to the potential energy of the $1/r_{ij}^{2}$ coupling.
In short, the former reads for a $\gamma =5/3$ fluid with density profile \Eq{defpressprof0}:
\begin{equation}
V_{2}=-\int_{0}^{r_{\rm m}} \frac{\kappa\gamma}{\gamma -1} \rho^{2/3} 4\pi r^{2} \rho \d r=-\frac{25\pi^{2} }{128}\kappa \rho_{0}^{5/3} r_{\rm m}^{3} \,,\EQN{def1}
\end{equation}
while the latter reads:
\begin{equation}
-\frac{K_{2}}{2}\iiint_{0}^{r_{\rm m}} \!\!
 \frac{8 \pi r^{2} \rho(r) \pi r^{'2} \rho(r')\d \mu \,\d r\, \d r' }{r^{2}+r^{'2}-2\mu\,r\, r'}= -\frac{9\pi^{4}}{320} \rho_{0}^{2} r_{\rm m}^{4 } K_{2}\,.
 \EQN{def2}
 \end{equation}
 Identifying \Eq{def1} and \Ep{def2} gives  (using $M=\pi^{2}\, \rho_{0} r_{\rm m}^{3}/8 $
 consistant with the density profile, \Eq{defpressprof})
 \begin{equation}
 \kappa= \frac{36}{125} \pi^{4/3}  M^{1/3} K_{2}.
\end{equation}
Similarly, requiring that $V_{2}$ and $V_{-2}$ balance at static equilibrium,
where
\begin{equation}
V_{-2}=\int \rho(r) \, \frac{ \omega^{2}\, r^{2}}{2} 4 \pi r^{2 }\d r=\frac{3\pi^{2}}{128} \omega^{2} \rho_{0} r_{\rm m}^{5}
\,, \EQN{defV-2}
\end{equation}
  yields
\begin{equation}
\rho_{0}=%\frac{1}{\pi}\left(\frac{2}{5}\right)^{3/2} 3^{3/4} \kappa^{-3/4} G^{\star 3/4} M^{5/4}
\left(\frac{5 G^{\star}}{3 K_{2}}\right)^{3/4}\, \frac{M}{\pi^{2}}
\,, \EQN{defrho0}
\quad {\rm and}
\end{equation}
\begin{equation}
r_{\rm m}=%\frac{10}{3^{1/2} \pi^{2/3}} \frac{\kappa^{1/2}}{M^{1/6} G^{\star 1/2} } 
2\left(\frac{3 K_{2}}{5 G^{\star} }\right)^{1/4}\, 
\,.\EQN{defrm}
\end{equation}
\Eqs{defpressprof}{defrm} allow us to relate the properties of the crystal to the properties of the 
analogous fluid system. 
Notice that $r_{\rm m }$ is independent of $M$ and $\rho_{0}$ is proportional to $M$.

\subsubsection{Figure of rotating  configurations}
%%%%%%%%%%%%%%
\label{s:secrot}
Equilibrium in the rotating frame with the angular  rate, $\M{\Omega}$ (not to be confused with $\omega$,
the strength of the harmonic potential introduced in \Eq{defW}),
 requires that:
\begin{equation}
\frac{1}{\rho}\nabla p=\nabla(\psi+\frac{1}{2}\Omega^{2}R^{2})\,,\quad{\rm where}\quad\psi=-\frac{1}{2}  G^{\star}M\r^{2}\,,
\EQN{eq-rot}
\end{equation}
where $R$ is the distance off axis ($r^{2}=R^{2}+z^{2}$).
Since $p=\kappa\rho^{5/3}$, it follows that, with $\Omega_{\star}^{2}=\Omega^{2}/(G^{\star} M)$,
\begin{equation}
\frac{5}{2}\kappa\rho^{2/3}=-\frac{G^{\star}M}{2}(z^{2}+(1-\Omega_{\star}^{2})R^{2})+{\rm const.}
\end{equation}
Let $\rho=0$ at $z=z_{0}$ on axis, then
\begin{equation}
\frac{\rho}{\rho_{0}}=\left(1-\frac{z^{2}}{z_{0}^{2}}-(1-\Omega_{\star}^{2})\frac{R^{2}}{z_{0}^{2}}\right)^{3/2}.
\EQN{def-rho-rotate}
\end{equation}
From this we can derive the mass, $M$ and the moment of inertia, ${\cal I}$ as
functions of the $z_{0}$, $\rho_{0}$  
\begin{equation}
M=\frac{\pi^{2}}{8}\frac{\rho_{0}^{}z_{0}^{3}}{1-\Omega_{\star}^{2}}\,,\quad{\rm and\quad{\cal I}=\frac{1}{4}MR_{0}^{2}} \,,
\end{equation}
where $R^{2}_{0}=z^{2}_{0}/(1-\Omega_{\star}^{2})$ and $z_{0}^{2}=5 \kappa /\rho_{0}^{2/3}/\mathcal{L}^{2} $.
\Eq{def-rho-rotate} generalizes \Eq{defpressprof} when the $\gamma=5/3$
fluid is given some angular momentum.
The ellipticity, $\varepsilon$, of the rotating configuration is given by
\begin{equation}
\varepsilon^{2}=1/(1-\Omega_{\star}^{2})=1/(1-\Omega_{}^{2}/\omega_{}^{2})\,.\EQN{ellip}
\end{equation}
\Fig{3dshellrot} displays the  corresponding configuration for $N=64$.

%%%%%%%%%%%%%%%%%%%%%%%%%%
\subsubsection{Dynamics of the   rotating oscillating $\gamma=5/3$ fluid}
%%%%%%%%%%%%%%%%%%%%

The basic pulsation of the $\gamma=5/3$ fluid in rotation is given by 
a time dependent uniform expansion/contraction  plus a rotation 
at constant angular momentum.
Thus 
\begin{equation}
\M{u}=\frac{\dot a}{a} \M{r}+ \frac{1}{a^{2}} {\tilde \M{J}}\times \M{r} \,,\EQN{defu}
\end{equation}
where   $a(t)$ is the ``expansion factor''
 of \Sec{nbody} and  $ {\tilde \M{J}}$ is the constant Lagrange multiplier associated with angular momentum conservation.
The acceleration involved in this motion are 
\begin{equation}
\frac{{\rm D}\V{u}}{{\rm D}t}=\frac{\partial \M{u}}{\partial t}+ \M{u}\cdot \nabla \M{u}= 
\frac{\partial \M{u}}{\partial t}+\nabla\left( \frac{1}{2} \M{u}^{2} \right) -\M{u}\times ( \nabla \times \M{u})\,.
\EQN{defudot}
\end{equation}
Given that $ \nabla \times \M{u}= 2 {\tilde \M{J}}/a^{2}$, putting \Eq{defu} into \Ep{defudot} yields
\begin{equation}
\frac{{\rm D}\V{u}}{{\rm D}t}=\nabla \left( \frac{\ddot a}{a} \frac{r^{2}}{2} -\frac{1}{2 a^{4}}({\tilde \M{J}}\times \M{r})^{2} 
\right)\,.
\EQN{defudot2}
\end{equation}
Differentiating  \Eq{defL} yields:
\begin{equation}
\ddot{a}=\frac{{\cal L}^{2}}{a^{3}}-\omega^{2}a\,, \EQN{addot}
\end{equation}
while Euler's equation reads:
 \begin{equation}
\frac{{\rm D}\V{u}}{{\rm D}t}=-\nabla\left(\frac{5}{2}\kappa\rho^{2/3}+\frac{1}{2}\omega^{2}r^{2}\right)\,.
\EQN{euler}
\end{equation}
Equating \Eq{euler} and \Ep{defudot2} together with \Eq{addot} yields:
\begin{equation}
 {\nabla}\left(\frac{5}{2}\kappa{\rho}^{2/3}+\frac{1}{2 a^{4}}{\cal L}^{2}{r}^{2}-
 \frac{1}{2 a^{4}}\left({\tilde \M{J}}\times \M{r}\right)^{2}\right)=0\,. \EQN{nabla}
 \end{equation}
Now in the rescaled space, $\rho \equiv {\tilde \rho}/a^{3}$, $\M{r} \equiv {\tilde \M{r}} a^{}$ and
  $\nabla \equiv {\tilde \nabla}/a^{}$, \Eq{nabla}  reads
\begin{equation}
 {\tilde \nabla}\left(\frac{5}{2}\kappa\tilde{\rho}^{2/3}+\frac{1}{2}{\cal L}^{2}{\tilde r}^{2}-\frac{1}{2}\left({\tilde \M{J}}\times \M{\tilde r}\right)^{2}\right)=0\,. \EQN{eq-rot3}
 \end{equation}
%
%which is  the braketed quantity of  \Eq{eq-rot}, but with
%rescaled variables replacing  the original ones,  and
 which is the equilibrium condition
	but with the time-dependently rescaled variables replacing the
	original ones. % and $G^{\star} M\rightarrow {\cal L}^{2}$.
 Hence it follows that in the comoving rotating frame, the 
 $\gamma= 5/3$ fluid will stratify in the same manner as \Eq{def-rho-rotate}\footnote{
 which  follows from integrating  \Eq{eq-rot3} while evaluating the integration 
 constant at ${\tilde \M{r} }=0$; this integration constant could in principle depend on time
 but because $\int {\tilde \rho} \d{}^{3} {\tilde r}\equiv M$  is independent of time,
  $5 \kappa {\tilde \rho}_{0}^{2/3}$ is indeed constant } but in the rescaled variables
 (${\tilde R_{0}}^{2}={\tilde z_{0}}^{2}/(1- {\tilde J}^{2}/\mathcal{L}^{2})$ and 
  ${\tilde z}_{0}^{2}=5 \kappa /{\tilde \rho}_{0}^{2/3}/\mathcal{L}^{2} $). This 
 maintains a {\sl constant} ellipticity during the oscillation, since 
  ${\tilde R_{0}}$ and ${\tilde z_{0}}$ are independent of $a(t)$.

%%%%%%%%%%%%%%%%%%%%%%%%%%%%%%%
\section{Applications: crystalline forms of pulsating equilibria}
\label{s:crystal}
%%%%%%%%%%%%%%%%%%%%%%%%%%%%%%%
%%%%%%%%%%%%%%%%%%%%%%%%%%%%%%%

Let us now study the discrete form of rotating pulsating 
equilibria that the systems described in \Sec{deriv} follow.
Let us first ask ourselves what would the final state of equilibrium in 
which $N$ particles obeying Eq.~(\ref{2}) would collapse to, if one adds 
a small drag force in order to damp the motions.

\begin{figure*}\unitlength=0.5cm
 {\centering \begin{tabular}{c c c }
\resizebox*{0.30000\textwidth}{0.30000\textwidth}{\includegraphics{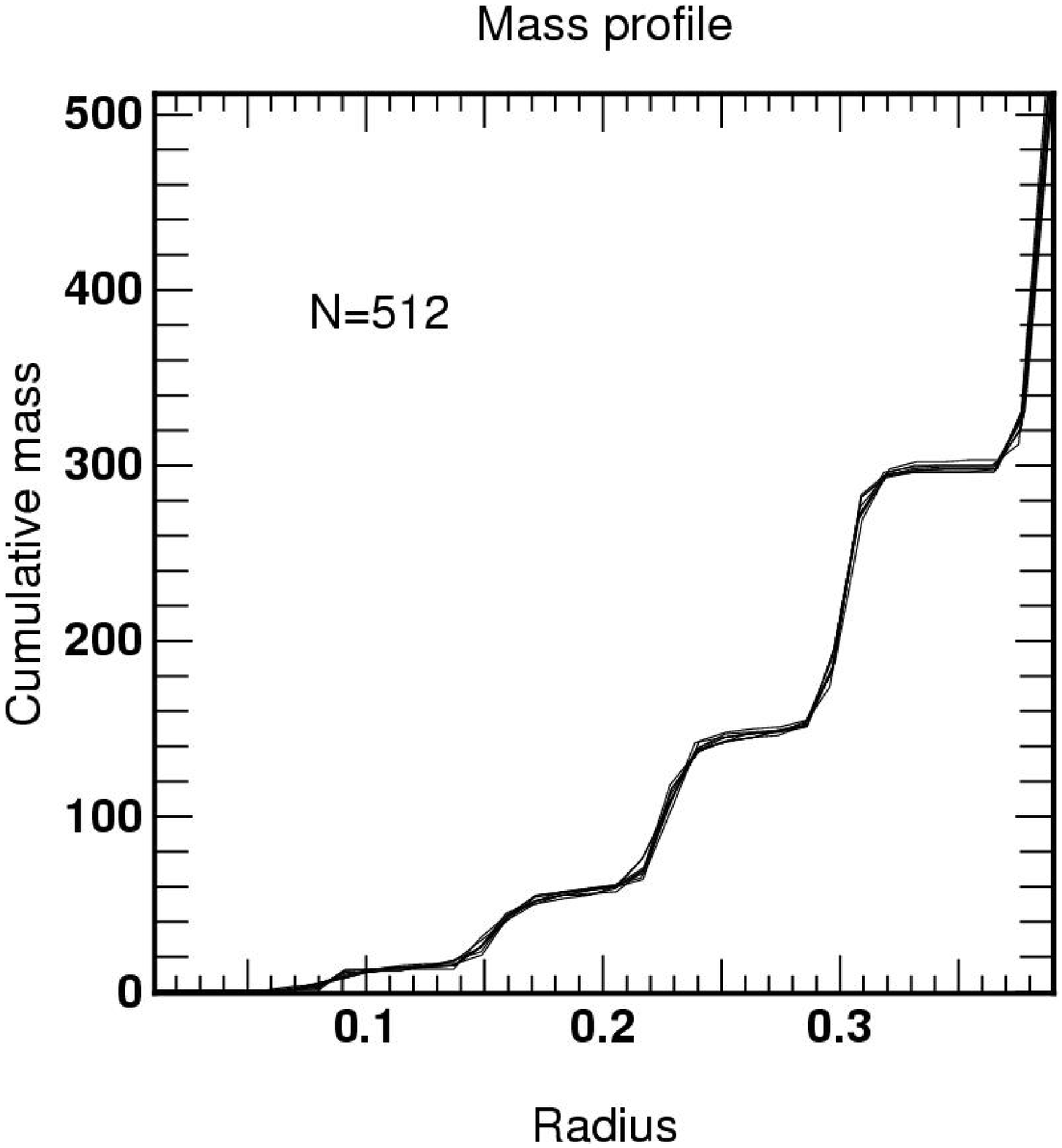}}  & 
\resizebox*{0.30000\textwidth}{0.30000\textwidth}{\includegraphics{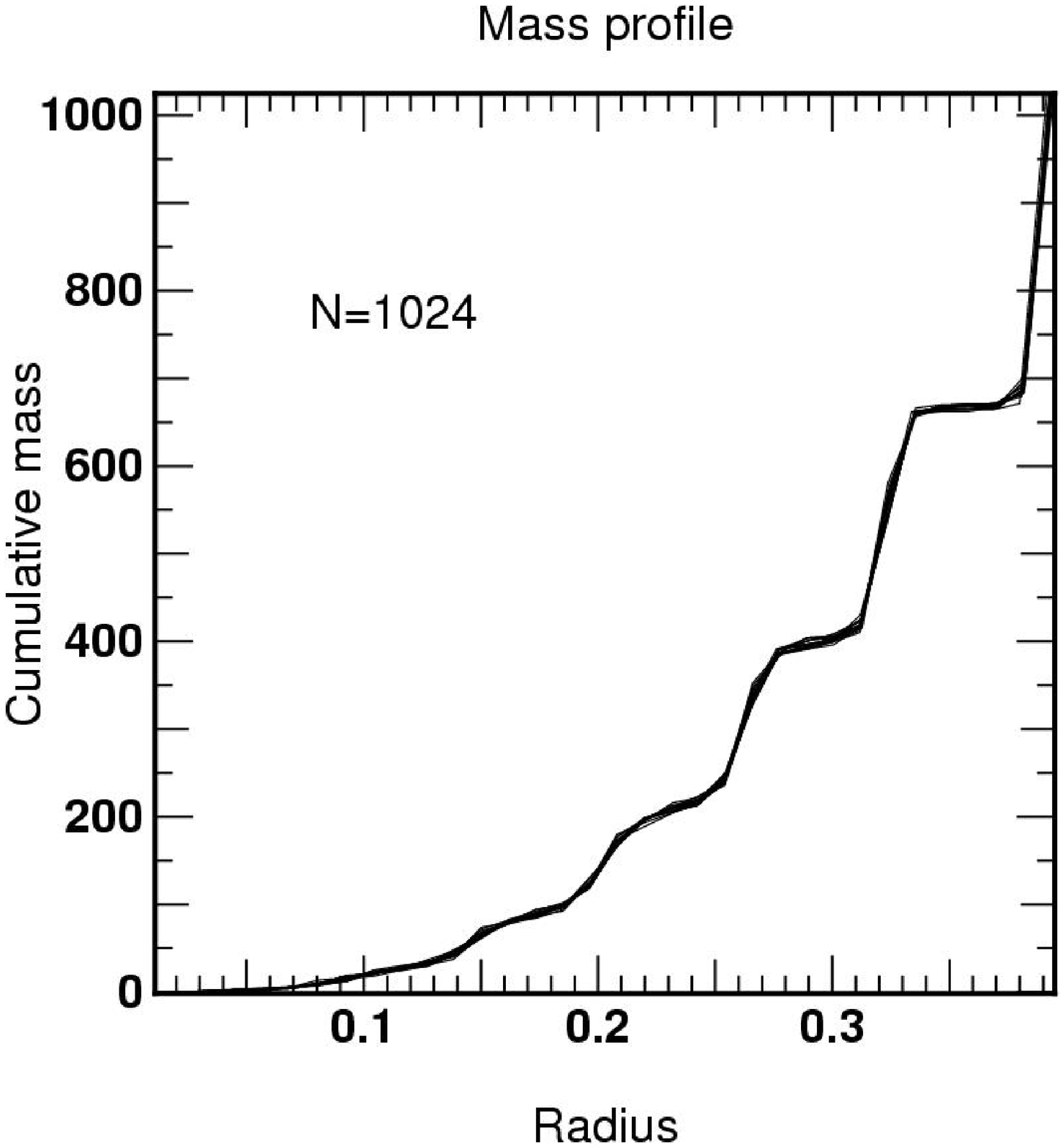}}  & 
\resizebox*{0.30000\textwidth}{0.30000\textwidth}{\includegraphics{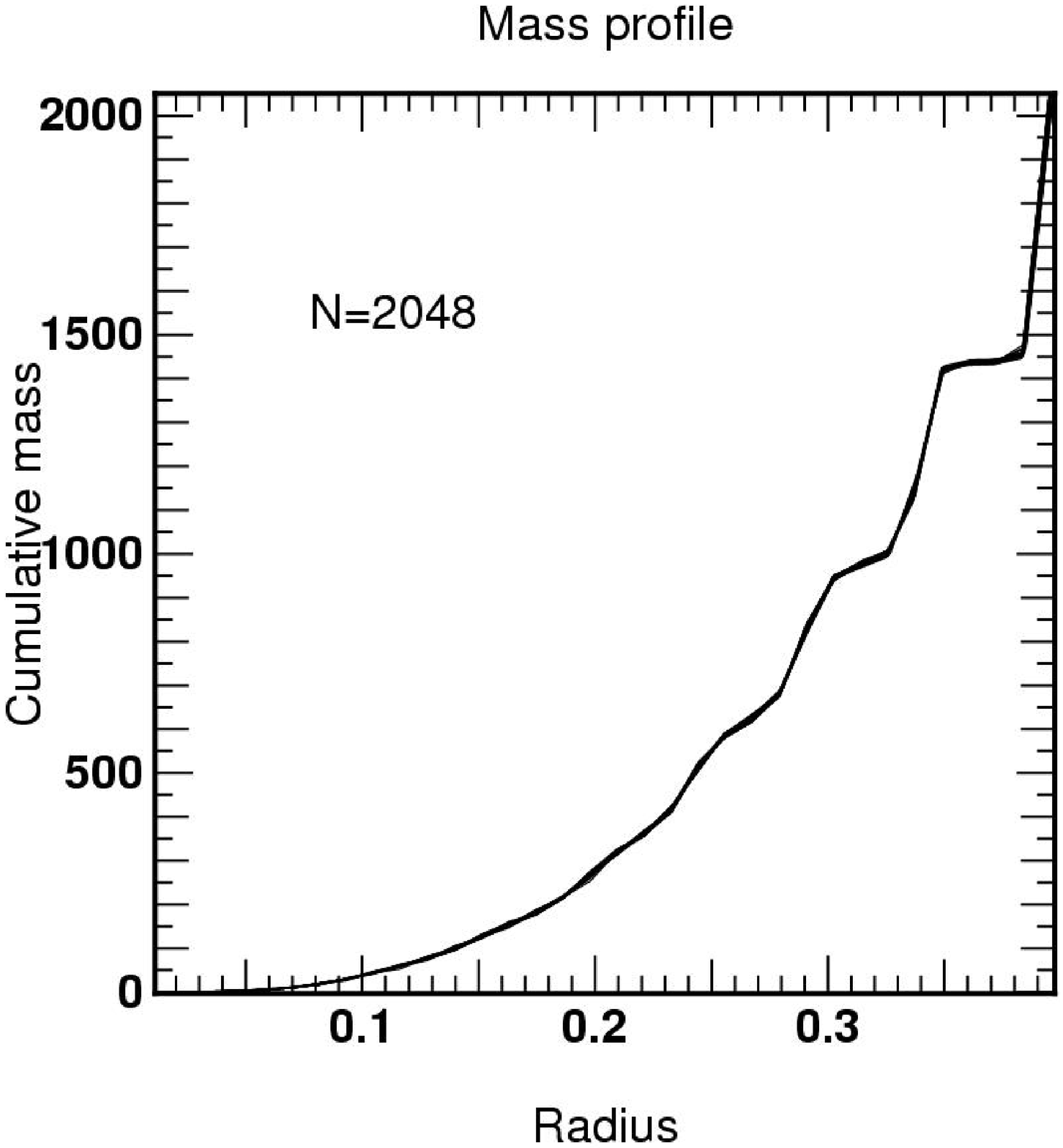}}  \\
\resizebox*{0.30000\textwidth}{0.30000\textwidth}{\includegraphics{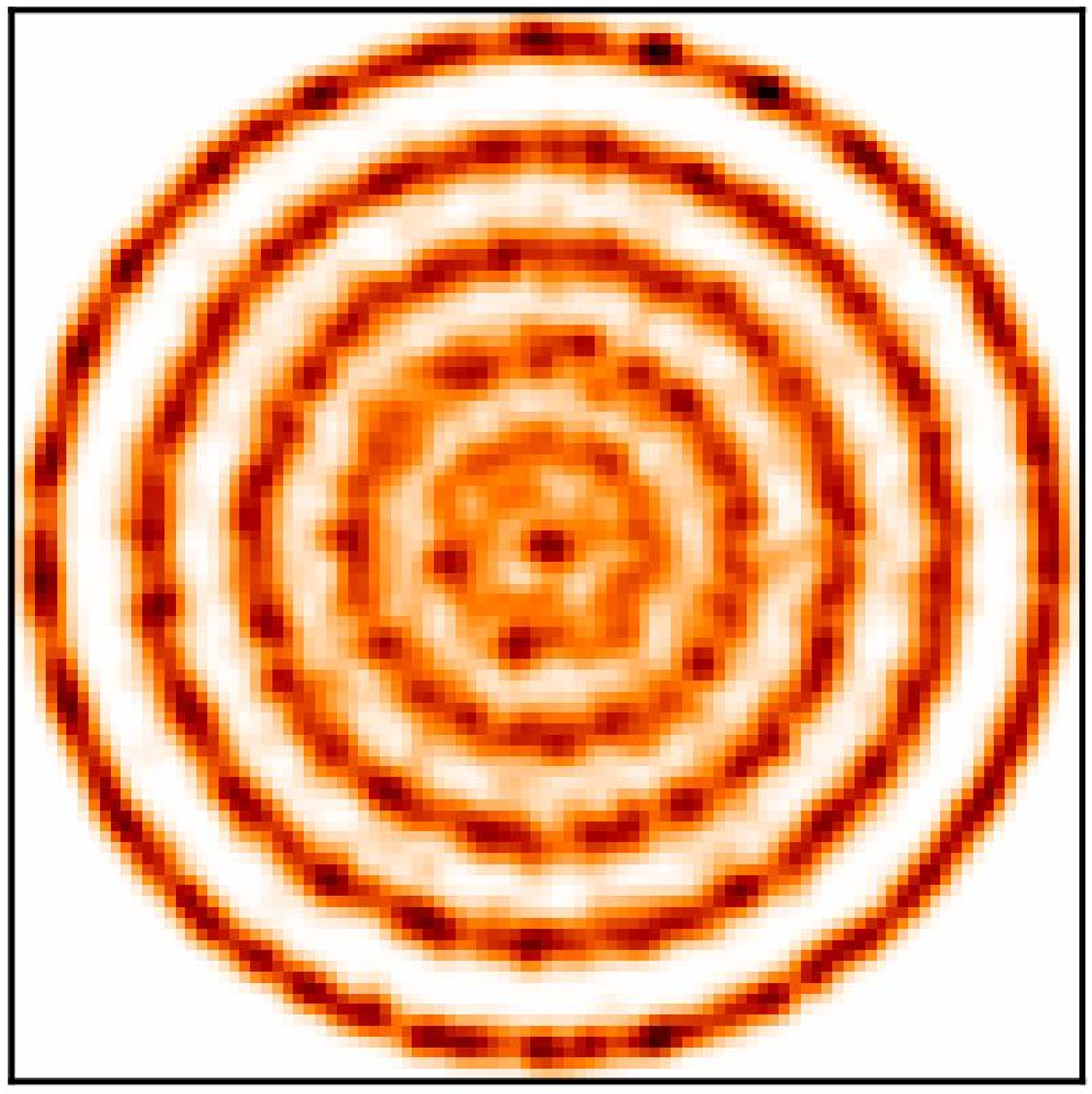}}  & 
\resizebox*{0.30000\textwidth}{0.30000\textwidth}{\includegraphics{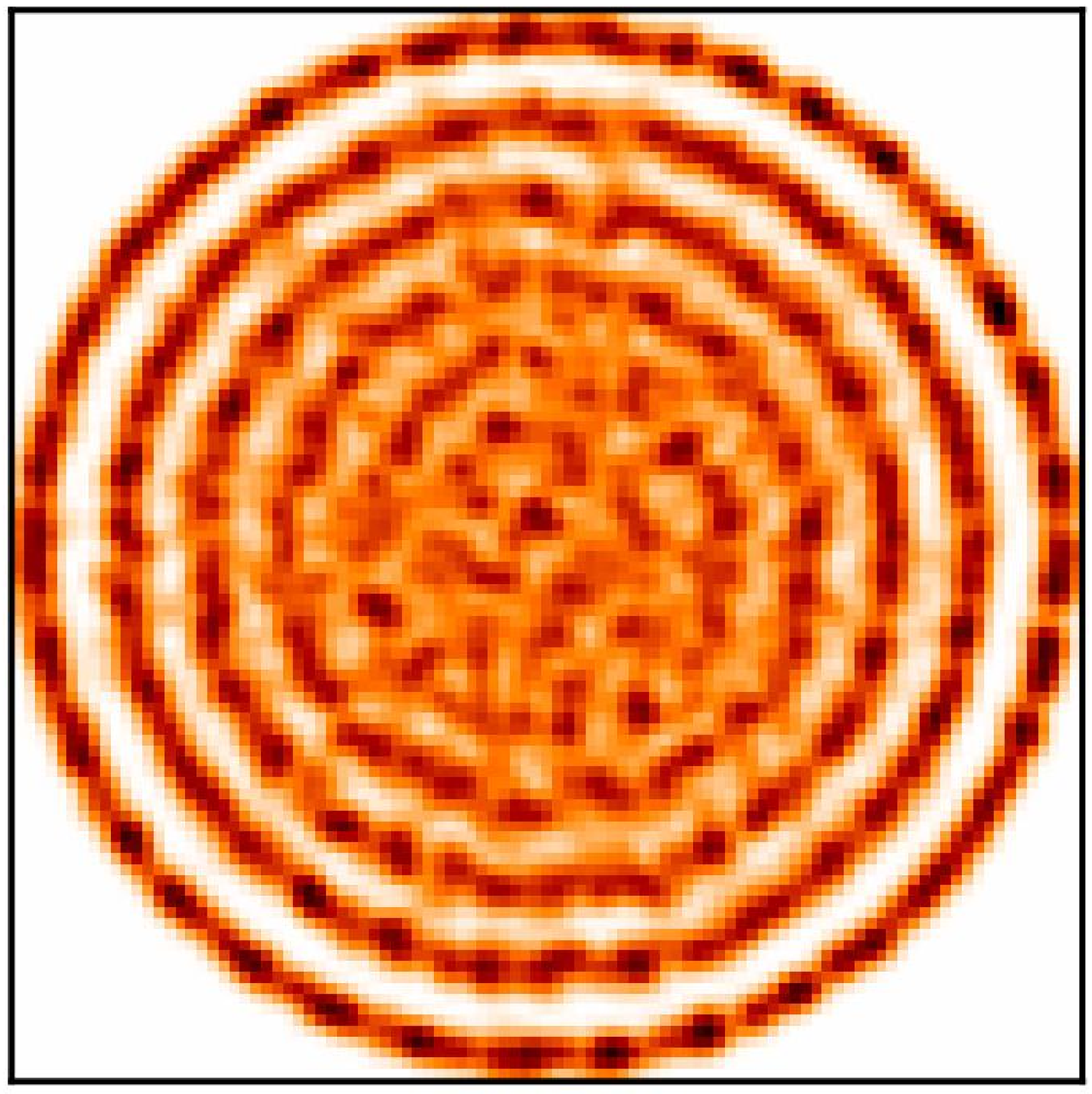}}  & 
\resizebox*{0.30000\textwidth}{0.30000\textwidth}{\includegraphics{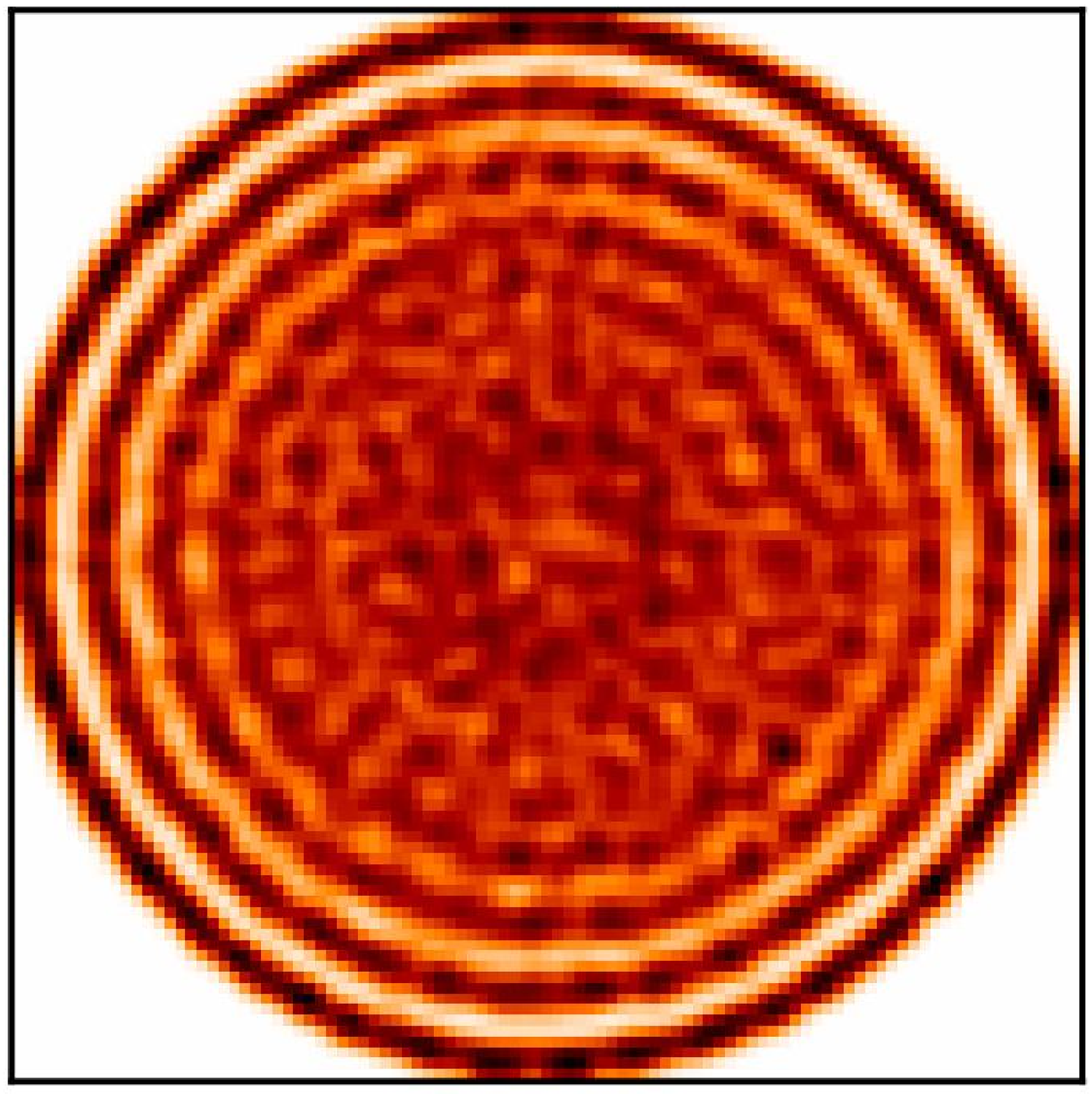}}   
 \end{tabular}\par}
\caption{{\sl top}: (Color 
  online)
Mean mass profile and section ({\sl top}) though a $N=512$, $N=1024$ and $N=2048$ crystal;
The mass profile is derived while stacking about 20 realisations of the lattice and binning the corresponding
density. The shell  structure ({\sl bottom}) is clearly apparent on these sections.
The number of shell, $N$, present is consistent with the prediction of \Sec{largeN}. The ordering of the inner regions is lower as $N$ increases 
since % the infall of the outer layer freezes the inner structure. 
the inner shells do not settle gently as they
	are disturbed by the infall of the outer layers.
% When rotation is introduced, as in \Sec{}, a similar spheroidal stratification occurs, and is
 % best seen when rescaling the particle positions along  the axis of rotation before stacking.
}
\label{f:profile-section}
\end{figure*}

\begin{figure}\unitlength=0.5cm
 {\centering \begin{tabular}{c  }
\resizebox*{0.350000\textwidth}{0.350000\textwidth}{\includegraphics{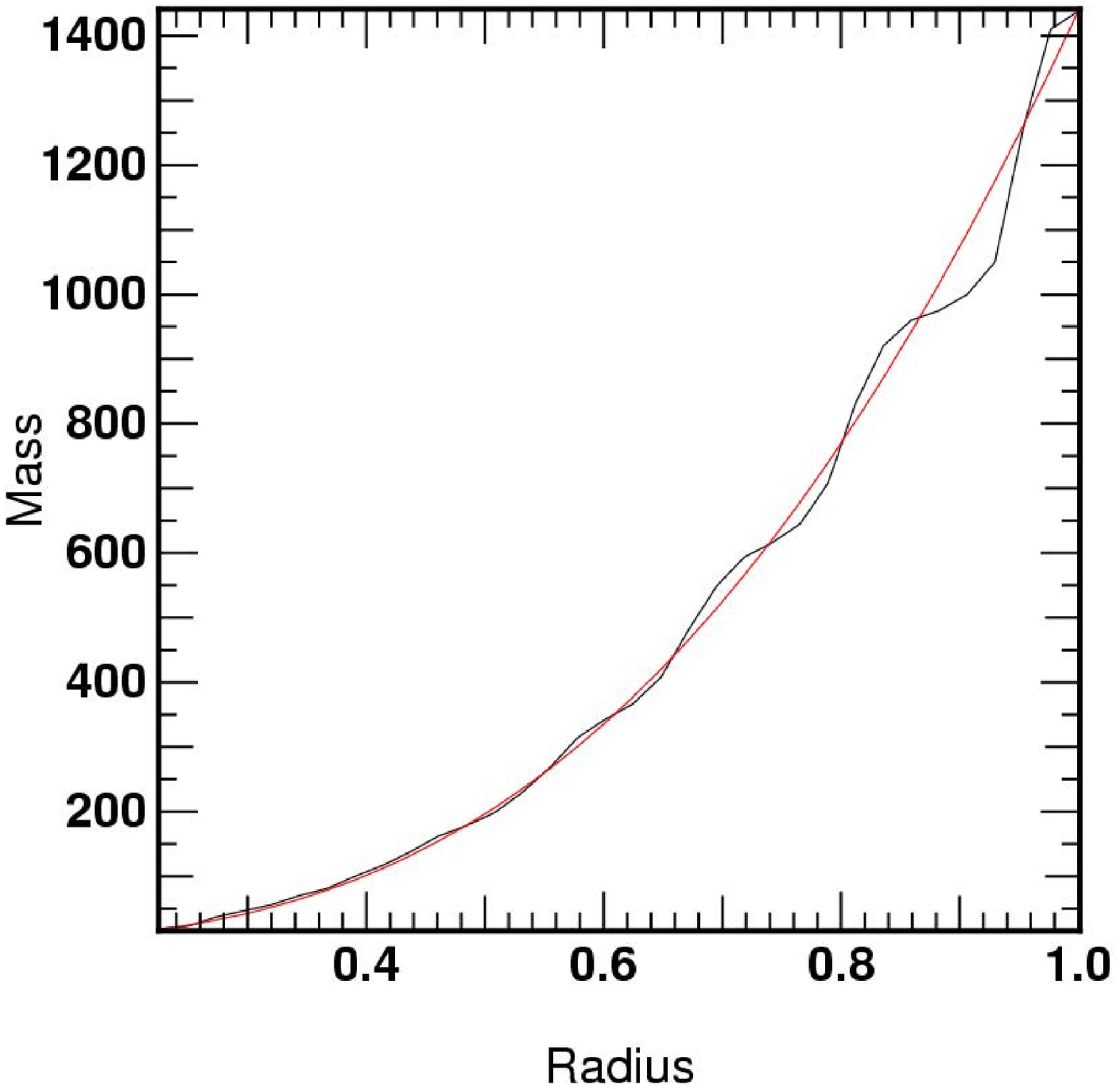}}  \\
\resizebox*{0.350000\textwidth}{0.350000\textwidth}{\includegraphics{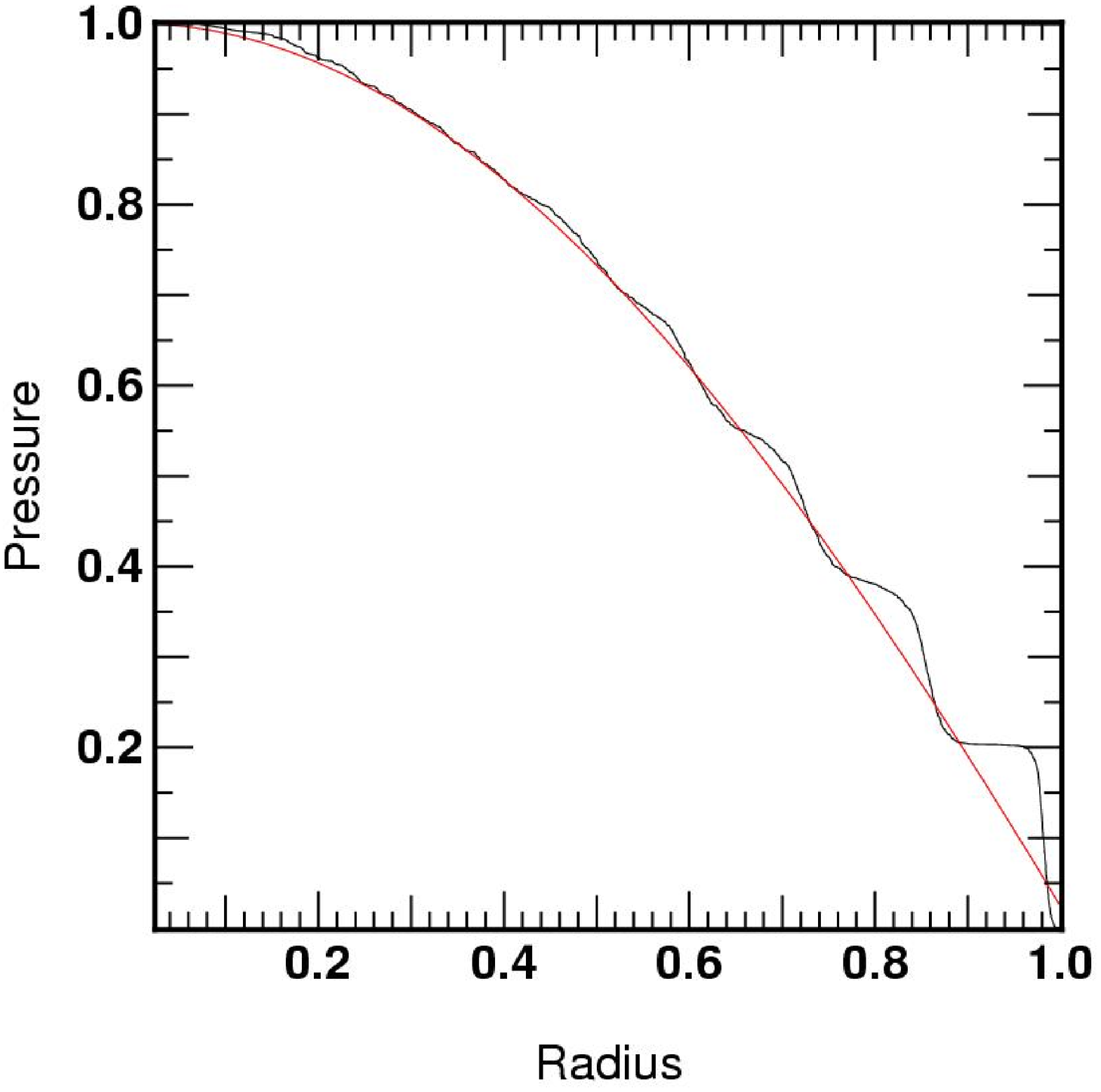}}  
 \end{tabular}\par}
\caption{ (Color 
  online)
  Mass profile and Pressure  of an $N=2048$ simulation together with {\em a fit of}
   its analytic prediction given by \Eq{defpressprof}  and \Ep{defmassprof} (see \Sec{stratification} and 
  \Fig{proffit}  for a discussion of 
the accuracy of this fit).
The mass profile is computed via the cumulative number of particles within
a given shell, while the pressure profile is estimated via
\Eq{defpn}.
} \label{f:figmassprof}
\end{figure}

\subsubsection*{ Numerical setup}

%%%%%%%%%%%%%%%%%%
A softening scale, $s$,  of $0.05$ was used\footnote{we also checked 
that our results remained the same with $s=0.01$}  so that The 
effective interaction potential reads
\begin{equation}
\psi_{12}=G^{\star} \frac{r_{12}^2}{2}+\frac{K_{2}}{r_{12}^2 +s^2}\,.
\end{equation} 
We may escale both the repulsive, $K_{2}$ and the attractive strength, $G^{\star}$ of the potential to one
by choosing appropriately the time units and the scale units.
As a check, we
 compute the total energy of the cluster, together with the 
invariant, $m {\cal L}^{2}/2$ and check its conservation.

%%%%%%%%%%%%%%%%%%%
\subsection{Static Equilibrium}
\label{s:static}
%%%%%%%%%%%%%%%%%%%

The relaxation towards the equilibrium should be relatively smooth in order to 
allow the system to collapse into a state  of minimal energy.
We will consider here two different damping forces; 
first (in \Sec{static}) an isotropic force, proportional to $-\alpha \M{v}$ so that both the 
rotation and the radial oscillations are damped;
or,  in \Sec{rotate},  a drag force
so that the  components of  the velocity are damped until  centrifugal equilibrium is reached.

 \subsubsection{Few particle  Equilibria}
%%%%%%%%%%%%%%%%%%%

Figure~\ref{f:fig-lown} displays the first few static equilibria, while 
Table~\ref{table} lists the first 18, which includes in particular the 
regular Icosahedron for $N=12$.
Strikingly,   roughly beyond this limit of about $N=20$,
there exists more than one set of equilibria for a given 
value of $N$, and the configuration of lowest energy is not 
necessarily the most symmetric. 
The overall structure is not far from hexagonal close packed, 
but with a spherical layering at large radii. 
Java animations describing the 
crystals are found at   {\em\tt http://www.iap.fr/users/pichon/nbody.html}.

\subsubsection{Larger N limit}
\label{s:largeN}
%%%%%%%%%%%%%%%%%%

Figure~\ref{f:profile-section} displays both 
the mean mass profile and the corresponding section though $512 \le N \le 2048$ lattices;
both  the mass profile and the sections are  derived while stacking different realisations of the lattice and binning the corresponding
density.  Note that the inner regions are more blurry as $N$ increases;
indeed, the inner layers are frozen early on by the infall of the outer layers.

Let  us estimate the pressure profile in our crystal.
In static equilibrium, given \Eq{defpressprof0}, we should have approximately
\begin{equation}
p(r)=\int \omega^{2}\rho(r) r \d r =\frac{ \omega^{2}}{4 \pi}
\int_{r}^{r_{\rm m}} \frac{\d m}{r}\approx \frac{\omega^{2}}{4\pi} \sum_{r'>r}\frac{m}{r'}\,. \EQN{defpn}
\end{equation}
\Eq{defpn} is compared to  \Eq{defpressprof0} in \Fig{figmassprof}.

\subsubsection{Mean density profile}
\label{s:mean-dens}
%%%%%%%%%%%%%%%%%%

\begin{figure}\unitlength=0.5cm
 {\centering
 \resizebox*{0.5000\textwidth}{0.50000\textwidth}{\includegraphics{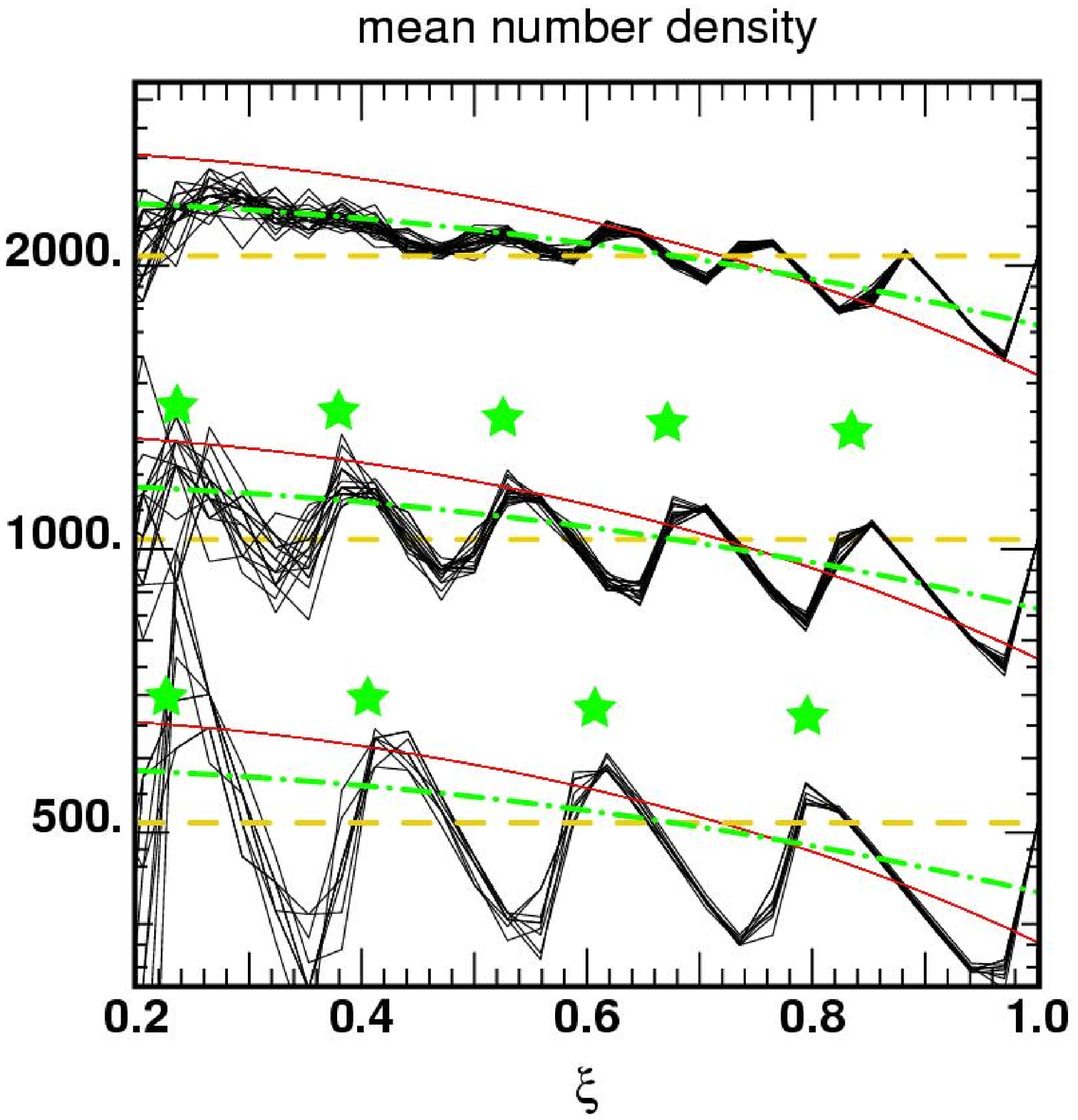}} 
 \par}
\caption{ (Color 
  online) shows the mean  number density profile, $N\,{\bar n}(\xi)$  for 
a set of about 10   $N=512$, $1024$ and $2048$ simulations
as a function of rescaled radius. The wiggles correspond to the shells seen in \Fig{profile-section}.
The thin line corresponds to the prediction of \Eq{defmassprof} for the fluid system, but with an outer
radius rescaled by the predicted radius, \Eq{defrm} divided by the measured outer radius, $r_{\rm m}$;
the dashed line corresponds to a uniform density solid, while the dotted dashed line
corresponds to the prediction  of \Eq{defrhom} (rescaled, see text). Clearly, this last model   gives the best fit to the mean profile, which suggests that 
for those values of $N$, the crystal does behave according to  the equilibrium \Eq{inteq}, and 
neither like a pure solid nor as a fluid. 
The stars correspond to the predictions of the shell radii given
by \Eq{defi}. 
}
\label{f:proffit}
\end{figure}

\Fig{proffit} shows the mean density within rescaled radius, $\xi=r/r_{\rm m}$ as a function of $\xi$ for three different simulations. The shell structure is very obvious 
but at larger $N$ values it is also clear that besides the oscillations due to shells the mean density falls off somewhat towards the outside.
However such a fall off is far less pronounced than that predicted by the polytropic model which gave ${\bar \rho}(\xi)= M(\xi)/(4 \pi r_{\rm m}^{3} \xi^{3/3})$
with $M(\xi)$ given by \Eq{defmassprof}. On further inspection, it transpired that this is due to the longer range of the repulsion force which is not correctly
 represented by the pressure analogy.  
 
 Consider a continuum density distribution with a repulsive force between elements $\d{m}$ and $\d{\bar m}$  derived from the potential $K_{2} \d{m} \d{\bar m}/|\M{r}-{\bar \M{r}}|^{2}$.
 Then, for a spherical density distribution, $\rho({\bar r})$, the total potential reads
 \[
 \psi(r)=K_{2}\!\!\int\!\!\!\!\int \frac{2\pi {\bar r}^{2} \rho({\bar r}) \d{\mu}\d {\bar r}}{r^{2}+{\bar r}^{2}-2\, r  {\bar r} \mu }= 2\pi K_{2}\!\!\int \frac{ {\bar r} \rho({\bar r})}{r }\log \left| \frac{r+{\bar r}}{r-{\bar r}}\right| \d{\bar r}
 \] 
The condition of equilibrium balances against the harmonic  attraction yields
\begin{equation}
-\frac{\d \psi }{\d r}= G^{\star} M r \,, \quad r\le r_{\rm m}\,.\EQN{inteq}
\end{equation}
Note that  {\em all} the mass, not just that inside radius $r$, contributes to the attraction for the linear law. 
\Eq{inteq} generates a linear integral equation for $\rho({\bar r})$.  A numerical solution to \Eq{inteq} shows that the mean density distribution agrees well with the simulations, once the 
shell structure is smoothed out (see \Fig{proffit}).  The solution to the integral equation is well approximated by 
\begin{equation}
\rho(r)=\rho_{0}\left(1-\frac{r^{2}}{6 r_{\rm m}^{2}} \right)^{3}\,,
\mbox{ where $r\le r_{\rm m}$ }\,, \EQN{defrho}
\end{equation}
and $0$ beyond.
The corresponding mean density within $r$, $\bar n$, scaled to one at the centre, reads 
\begin{equation}
{\bar n}(\xi)=\frac{3}{\rho_{0}\xi^{3}}\!\int^{\xi} \!\! \rho(r_{\rm m} \xi) \xi^{2} \d{\xi} =\left(1 - \frac{3\,\xi^2}{10} + \frac{\xi^4}{28} - \frac{\xi^6}{648}\right)\! , \EQN{defrhom}
\end{equation}
with $\xi\equiv r/r_{\rm m}$,
and  $n(r)\equiv\rho(r)/m$, (see \Fig{proffit})
%
%\begin{equation}{\bar n}(<\xi)\equiv  \frac{n_{0}}{(\xi^{3}/3)} \int_{0}^{\xi}(1-\xi^{2}/6)^{3} \xi^{2} \d \xi ,
%\end{equation}
so that at $\xi=1$, ${\bar n}(1)\equiv X=16\,651/22\, 680$. 
The simulations have been scaled so that $\xi$
is one at the outermost particle. That will not be at the point at which
the theoretical smooth density falls to zero which we have called $\xi =1$
in our theoretical calculation. In practice this falls outside the last
particle, hence we have to rescale the theory's $\xi$ by a factor $f$ which
is determined to make the theoretical mean density profile, $\bar n$, fit the
profile of the simulations' mean density. Once this scaling $f$ has been
determined the predicted mean density is given in terms of the observed $\xi$
by the expression
$X {\bar n} ( f \xi) N/[4/3 \pi(f r_{\rm m})^{3}]$.
%Hence we have to use a radial scale for the observation which reaches
% the mean density $X$ when $\xi$ reaches $1$ (which may well be 
% outside the last shell, depending on $N$). Thus our first step is to re-calibrate
% the radial scale so that the mean density fits the mean density of the simulations.
%

\subsubsection{Stratification in spherical shells}
\label{s:stratification}
%%%%%%%%%%%%%%%%%%

In atoms, the main gradient of the potential is towards the nucleus, 
so the shell structure is dominated by the inner shells which have
large changes in energy. In our systems, the global potential gradient is proportional to $r$ so the largest potential differences are on the outside. As a consequence, the system adjusts itself so that the outermost shell is 
almost full, and the central region is no longer shell dominated.
Instead of counting shells outwards from the middle as in atoms, it is best to
count shells  inwards from the outside where they are best defined.
For 
shells which  are numbered from the outside,
given that $\d r/n^{-1/3}(r)$ is the fractional increase
in layer number,
we have (using \Eq{defrho})
\begin{equation}
\int_{r_{i}}^{1} \frac{\d r}{n(r)^{-1/3}} = \frac{
n_{0}^{1/3} r_{\rm m}}{X}\left[
1-\xi_{i}-\frac{1}{18}(1-\xi_{i}^{3})
\right]=i \,,\EQN{defi} 
\end{equation}
 with $n_{0}\equiv \rho_{0}/m= 3N/(4 \pi r_{\rm m}^{3})$.
Conversely, the number of particles, $N(\le i)$ within (and including) layer $i$  is given 
by
\begin{equation}
N(\le i)= \frac{N}{X}\, \xi_{i}^{3} \! \left[ 
1-\frac{3}{10}\xi_{i}^{2}+\frac{1}{28}\xi_{i} ^{4}-\frac{1}{648}\xi_{i} ^{6}
\right]. \EQN{defNi}
\end{equation}
Solving  the implicit \Eq{defi} for the relative radius $\xi_{i}$ of layer $i$ and putting it into \Eq{defNi}
yields the number of particles within layer $i$ as a function its rank. 
The radii of the layers, $\xi_{i}  r_{\rm m} $ follows from inverting \Eq{defi} for 
$\xi_{i}$.
Examples of such shells are shown in  \Fig{3dshell} 
and \Fip{profile-section}, while the corresponding mass profile 
checked against \Eq{defmassprof} in \Fig{figmassprof}.
The number of shell present in \Fig{profile-section} is consistent with the prediction of
\Eq{defNi}.
\subsection{Rotating  Perpetually Pulsating Equilibria}
\label{s:RPPE}
%%%%%%%%%%%%%%%%%%%

%%%%%%%%%%%%%%%%%%%
\subsubsection{Rotating crystals}
\label{s:rotate}
\begin{figure}\unitlength=0.5cm
 {\centering
 \resizebox*{0.40000\textwidth}{0.300000\textwidth}{\includegraphics{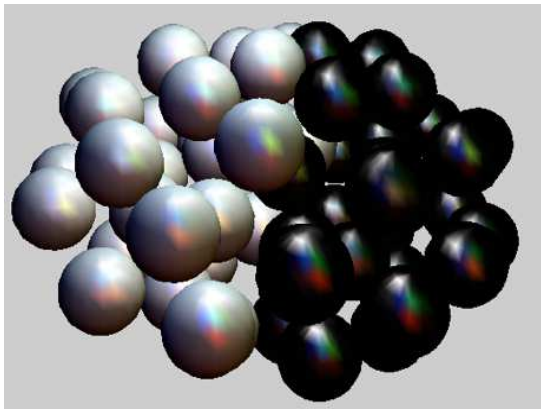}} 
 \par}
\caption{  (Color 
  online) spheroidal crystal corresponding 
to the final state of a run  of $N=64 $ with a damping force of the form
  $-\alpha \left( \M{v}- {\mathcal I}^{-1}\cdot {\tilde \M{J}}
\right)$.
Half of the particles was painted one color, while the other half was left some other color.
 The shape preserving oscillation is achieved by 
rescaling all position by some factor, $\lambda$, and rescaling all velocities 
by the factor $1/\lambda$. If the rescaling does not preserve momentum of each particle, 
some of the excess energy may go into heating the system and breaking its structure,
which will induce mixing of the two populations (see
\Sec{relax}  and  paper III). 
Java animations corresponding to the  corresponding oscillating rotating 
crystals are found at   {\em\tt http://www.iap.fr/users/pichon/nbody.html}. 
}
\label{f:3dshellrot}
\end{figure}

%%%%%%%%%%%%%%%%%%%
The rotating crystal is achieved in steps; 
first, for each particle in the lattice, we 
added a velocity kick so that $\M{v}\rightarrow \M{v}+\M{\Omega} \times \br$.
We then rescale the $z$ 
coordinate and 
the $v_z$ (as defined along the momentum direction) of each particule by a
constant factor.
The system has then the ability to redistribute it momentum along the 
oscillating particles.
We then add a drag force\footnote{which can be shown to be truly dissipating energy} 
proportional to   $-\alpha \left( \M{v}- {\mathcal I}^{-1}\cdot {\tilde \M{J}}
\right)$, 
so that the  components of  the velocity are damped until  centrifugal equilibrium is reached.
This defines the rotating spheroid equilibrium. 
An example of such a configuration is shown in \Fig{3dshellrot} for $N=64$.
The properties of the corresponding analogous fluid system are 
described in \Sec{secrot}.

%%%%%%%%%%%%%%%%%%%
\subsubsection{Rotating pulsating crystals}
\label{s:rotpulse}
%%%%%%%%%%%%%%%%%%%

In order to create a pulsating configuration which preserves 
the shape of the rotating spheroid, we rescaled all the  positions by some factor, $\lambda$, and rescaled accordingly all velocities 
by the factor $1/\lambda$, so that  angular momentum  is preserved.
When this special condition is met, the system oscillates and pulsates  without {\sl any}
form of relaxation ({\it cf.} \Sec{thermo}).
%, even though the force field does not  impose such a strong invariance.  
Note that this situation differs from normal modes of more 
classical systems, which 
do preserve  the shape of a given oscillation, but 
might not
involve the same particles at all times.
Java animations describing  the pulsating and rotating 
crystals are found at  {\em\tt http://www.iap.fr/users/pichon/nbody.html}.
%\Xtophe{think of some figure ??}

\begin{figure}\unitlength=0.5cm
 {\centering
 \resizebox*{0.350000\textwidth}{0.350000\textwidth}{\includegraphics{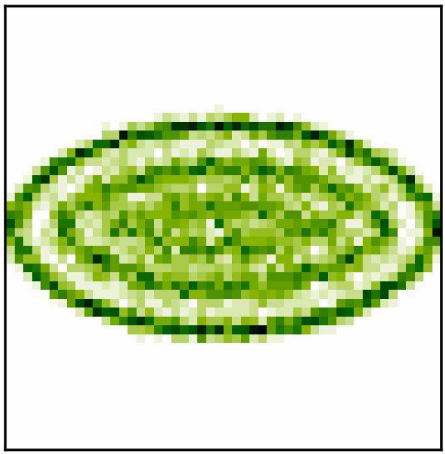}} 
 \par}
\caption{ (Color 
  online) 
Same as \Fig{profile-section}, but for a set of  20 rotating configuration
of $N=256$ particles  launched according to the prescription 
given in \Sec{rotate}.  The shell  structure is also clearly apparent on this section.
The ellipticity of the crystal is found to be in agreement with 
\Eq{ellip}.
}
\label{f:fig-ellip}
\end{figure}

%%%%%%%%%%%%%%%%%%%
\subsection{Thermodynamics of dissolving crystal}
\label{s:thermo}
%%%%%%%%%%%%%%%%%%%
%%%%%%%%%%%%%%%%%%%

 \begin{figure} 
\centering
\resizebox{0.45\textwidth}{0.45\textwidth}{\includegraphics{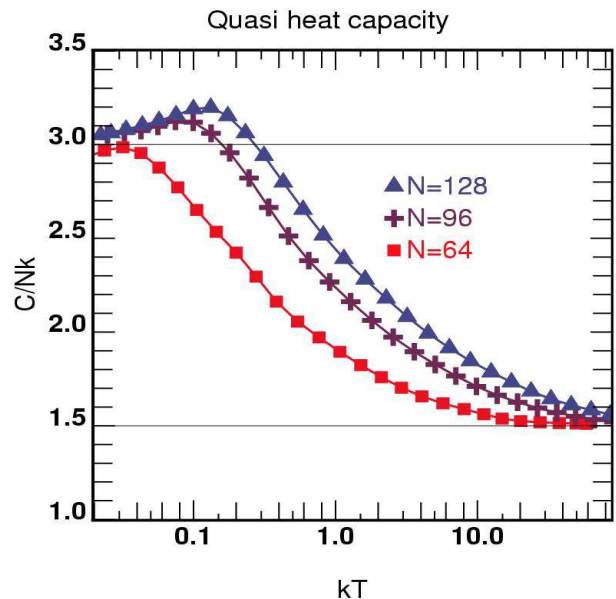}}
\caption{ (Color 
  online) quasi
specific heat measured by increase of quasi energy with quasi temperature in units of $k N$; 
the quasi temperature is  $a^{2}$ times the kinetic energy relative 
to the time dependent ``Hubble flow'' 
 divided by $ 3/2 \, N k$;  the quasi energy is  the sum of the quasi kinetic 
 energy plus $W_{2}$;  $W_{2}$ is 
 $a^{2}$ times the repulsive  part of the potential energy.
  During the large amplitude pulsation of the system,
  the quasi energy is conserved. It is the quasi energy that is shared between the different components in the statistical equilibrium. 
 Since $W_{2}$ is only repulsive, the system displays characteristics similar to 
 a hard sphere fluid. 
 %Hence there is no latent heat as there are no bonds  
 %and therefore no bonds to break.
 There is a continuous transition between the cold crystal lattice and the ``free''  fluid. Each point is derived over a set of about 20 independent runs.  
 The initial condition is set up with some random motions above the 
 lattice equilibrium.
% \Xtophe{bump induced by softening?}
  }
\label{f:heat} 
\end{figure}

\subsubsection{Specific heat \& evaporation}
%%%%%%%%%%%%%%%%%%%

How does the shell structure disappear as a function of temperature increase ?
Fig.~\ref{f:heat} displays the quasi-specific heat measured by the  increase of quasi energy with quasi temperature in units of $k N$;  Here
the quasi temperature is  $a^{2}$ times the kinetic energy relative to the time dependent  ``Hubble flow'' 
 divided by $ 3/2 \,N k$;  while the quasi energy is  the sum of the quasi kinetic energy plus $W_{2}$ given by \Eq{defW2}. During the large amplitude pulsation of the system,  the quasi energy is shared between the different components in the statistical equilibrium. 
 Since $W_{2}$ is only repulsive, the system displays characteristics similar to 
 a hard sphere fluid. 
 %Hence there is no latent heat as there are no bonds  
 %and therefore no bonds to break.
The quasi-heat capacity changes from the solid's $3N k$ to the
	gaseous $3/2\, N k$ as expected.

\subsubsection{Relaxation of the rotating pulsating configuration}
\label{s:relax}
%%%%%%%%%%%%%%%%%%%

\begin{figure}\unitlength=0.5cm
\centering
\resizebox{0.45\textwidth}{0.45\textwidth}{\includegraphics{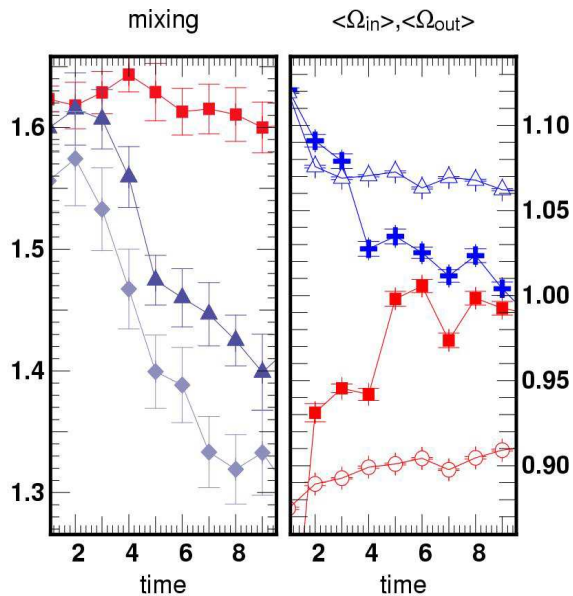}}
 \caption{ (Color 
  online) {\sl left panel}:
$ \Delta n_{\rm W B}/ { \bar  n_{\rm W B} } $,
the relative dispersion in mixing as  a function of time for a set of 
$N=64$ particles in a situation where the inner  fraction of particles 
was rescaled along the rotation axis in order to induce some momentum 
exchange.  Here three scalings (1.2, 1.6 and 2) were imposed (from top to bottom), corresponding 
to an increasing heat content which induce a faster relaxation.
%The characteristic mixing time-scale,  $\Delta \tau$ is also shown here. 
The initial condition corresponds to the prescription 
described in  \Sec{rotate} and  \Fig{3dshellrot}.
 {\sl Right panel}:
%the corresponding rate at which the system will 
%redistribute momentum: 
$
\Omega_{\rm in}(t)/{\bar \Omega}(t)$  and 
$\Omega_{\rm out}(t)/{\bar \Omega}(t)$  as 
a function of time;  again the hotter initial configuration (rescaled by 1.8 compared to 1.2)
(full symbol) reaches a 
regime of uniform rotation more rapidly.
 % {\sl Bottom panel}: the thermalisation of the velocities in the 
% comoving frame. The velocity distribution of the 
% $R$ and $z$ components of  $ \M{u}_{\rm pec}/a$ after ? oscillations of 
%a crystal launched as before. 
%\Xtophe{is le theta component going to be different ?}
} \label{f:relax}
\end{figure}

Let us split the particles within our  spheroidal equilibrium in two sets
as a function of axial radius, $R$, one corresponding to an inner ring, and one corresponding 
to an outer ring (which should initially rotate at the same angular rate)
 and rescale the $z$ component of each inner particle 
by a factor of $2$ so that it does not satisfy the equilibrium condition. 

A measure of the thermalisation  is given by the rate at which the system will 
redistribute momentum in order to achieve uniform rotation again.
\Fig{relax} ({\em right panel}) represents $
\Omega_{\rm in}(t)/{\bar \Omega}(t)$  and 
$\Omega_{\rm out}(t)/{\bar \Omega}(t)$  
as a function of time (with ${\bar \Omega}(t)=[\Omega_{\rm in}(t)+
\Omega_{\rm out}(t)]/2$).

%In the spirit of paper III, we compute the distribution of velocities 
%in the comoving rotating pulsating frame.  Let $\M{v}_{\rm pec}$ be the velocities of 
%particles w.r.t. that frame:
%\begin{equation}
%\M{v}_{\rm pec} \equiv \M{v}-\frac{\dot a}{a} \M{r}- \frac{1}{a^{2}} {\tilde \M{J}}\times \M{r} \,.\EQN{defdu}
%\end{equation}
%\Xtophe{do we want to add this ?}
%\Fig{relax} (bottom panel)  shows the velocity distribution of the 
%radial  (resp, vertical) component of $\M{u}_{\rm pec}$ after ? oscillations 
%of a crystal made of $N=?$ particles. 
%\Xtophe{check its isotropic in $ \M{u}_{\rm pec}$???} 

\subsubsection{ Mixing  of the rotating pulsating configuration}
%%%%%%%%%%%%%%%%%%%

A measure of mixing  is given by the rate at which the system
becomes uniform, when 
it is started as two distinct phases. Let us start by a configuration where half of the particles
on one side of the spheroidal rotating equilibrium are colored in WHITE, 
and the other half, in BLACK, and rescale again the coordinate along 
the rotation axis by a factor of two.  The 
redistribution of the excess energy in height will 
convert a fraction of it into heat. 
\Fig{relax}  ({\em left panel}) represents $ \Delta n_{\rm W B}/ { \bar  n_{\rm W B} } \equiv
 {\rm RMS}( n_{\rm B}(t)-n_{\rm W}(t))/
{\rm RMS }( n_{\rm B}(t)+n_{\rm W}(t) )$, with $ n_{\rm  R}$ the density 
of WHITE particles, and $ n_{\rm  B}$ the density 
of BLACK particles
as a function of time.
%Initially, we expect
% $\langle| n_{\rm B}(t)-n_{\rm W}(t) |\rangle=
%  \langle| n_{\rm B}(0)| \rangle =\langle| n_{\rm W}(0) |\rangle$, so that
%   $ \Delta n_{\rm W B}/ { \bar  n_{\rm W B} }=1$,
%whereas at later times $ n_{\rm B} \rightarrow n_{\rm W}$.
%Let us call $\Delta \tau$ the time at which $ \Delta n_{\rm W B}/ { \bar  n_{\rm W B} }$ reaches $1/2$
% \Fig{relax}  (bottom panel)  shows the evolution 
%of $\Delta  \tau$ versus $\Omega$ and $T$, the temperature of the 
%rotating crystal.
%\Xtophe{an alternative is to define an entropy
%of 
%$S_{A+B}-S_{A}-S_{B}$ ?}
In practice, we bin the $x$--$y$ plane in the comoving coordinates and 
estimate   $ \Delta n_{\rm W B}/ { \bar  n_{\rm W B} }$ accordingly.

%%%%%%%%%%%%%%%%%%%%%%%%%%%%%%
\section{Conclusion}
%%%%%%%%%%%%%%%%%%%%%%%%%%%%%%
%%%%%%%%%%%%%%%%%%%%%%%%%%%%%%

%\begin{figure}\unitlength=0.5cm
%\vskip 5cm
% \caption{ {\sl left panel}:
%$ \Delta n_{\rm W B}/ { \bar  n_{\rm W B} } $,
%the relative dispersion in mixing as  a function of time.
%The characteristic mixing time-scale,  $\Delta \tau$ is also shown here. 
%The initial condition corresponds to the prescription 
%described in  \Sec{rotate} and  \Fig{3dshellrot}.
% {\sl Bottom panel}:  Evolution of $\Delta \tau$ as a function of 
%angular momentum ${\tilde \M{J}}$ 
% and temperature $T$. The rotation slows down mixing since particles 
% spend a larger fraction of their time without interacting 
% with each other. \Xtophe{verifier}
%} \label{f:relax2}
%\end{figure}

In this paper, we have  demonstrated that 
 very special systems, those for which the 
oscillation of the system separates off dynamically from the beheaviour of the 
rescaled variables, can form (possibly spinning) solids
(albeit ones that pulsate in scale). We studied their phase transition
and the corresponding ``specific heat'' as these structures 
melt. As expected,  we found that  the heat capacity halve to that of 
a free fluid, but the phase transition appears to occur gradually 
even at large $N$. 
Although we expected a set or regular and semi regular solids,
which we did find for small $N$, and an hexagonal close packing 
of most of the system at larger $N$, we did not foresee the strong
shell structure found. 
Nevertheless, we constructed a theory that explained this 
shell structure once it had been recognized. 
We also investigated the evolution of the lattice, as some rotation 
was imposed on the structure and studied its relaxation under such 
circumstances, both in terms of mixing and for  the redistribution of angular momentum. 
Another by-product of our study was that the internal energy of 
the corresponding barotropic fluid scales as $\rho^{\gamma -1}$, 
{\it i.e.} as $({\rm scale})^{-3(\gamma-1)}$ so that for $\gamma=5/3$,
it scales like $({\rm scale})^{{-2}}$. 
Thus our investigation applies 
equally to a $\gamma^{5/3}$  fluid with a quasi gravity such  that every 
element of fluid attract every other in proportion to their separation.
It is well known, since Newton, that such attractions are equivalent 
to every particle being attracted in proportion to its distance to 
the centre of mass as though the total mass were concentrated there.
With such a law of interaction, we described the possible set of 
rotating pulsating equilibria it may reach, and found them
 is qualitative agreement with our simulations of the corresponding crystals,
 though the details of their mass profile differed:
 this fluid model failed to give the correct layering,
while a more acurate model gives it correctly.
Recently \cite{monaghan} used simulations of such systems to
test their smooth particle hydrodynamics codes.
%without knowledge of 
%its connection to the N-body problem discussed here
%\Xtophe{Donald are you sure about this ?
%It doesn't matter, but you did mention these systems to 
%monaghan at the RAS where both him and I were your guest}.

\vskip 1cm

{\bf Acknowledgements}
{\sl
We thank David Wales for
verifying our structures with his more thorough and well tested program.
We  would  like  to  thank D.~Munro  for  freely
   distributing   his  Yorick  programming   language and opengl interface  (available   at  {\em\tt
  http://yorick.sourceforge.net}) which we used to implement
  our N-body program. D.~LB acknowledges support from EARA 
  while working at the Institut d'Astrophysique de Paris where 
 this work was completed. 
}

\def\BIBITEM#1#2#3{\bibitem[\protect\citename{#2, }#3]{#1}}

\def\aj {Astronomical Journal}
\def\apj{ApJ }
\def\apjs{ApJS }
\def\pasj{Publ. Astron. Soc. Japan}
\def\aj{AJ }
\def\mn{MNRAS }
\def\pasp{PASP }
\def\apjl{ApJL \rm}
\def\mnras{Mon. Not. R. Astron. Soc}
\def\aeta{A\&A }
\def\aetal{A\&AL }

{}

\appendix

%%%%%%%%%%%%%%%%%%%%%%%%%%%%%%%%%%%%%%%%%
%%%%%%%%%%%%%%%%%%%%%%%%%%%%%%%%%%%%%%%%%
%%%%%%%%%%%%%%%%%%%%%%%%%%%%%%%%%%%%%%%%%
%%%%%%%%%%%%%%%%%%%%%%%%%%%%%%%%%%%%%%%%%
%%%%%%%%%%%%%%%%%%%%%%%%%%%%%%%%%%%%%%%%%
%%%%%%%%%%%%%%%%%%%%%%%%%%%%%%%%%%%%%%%%%
%%%%%%%%%%%%%%%%%%%%%%%%%%%%%%%%%%%%%%%%%

\end{document}